\documentclass[]{aa}
\usepackage[dvipsnames]{xcolor}
\usepackage{soul}
\usepackage{ulem}
\usepackage{graphicx}
\usepackage[varg]{txfonts}
\usepackage[colorlinks,citecolor=blue,urlcolor=blue]{hyperref}
\usepackage[flushleft]{threeparttable}
\usepackage{natbib}
\usepackage{verbatim}
\usepackage{float}
\usepackage{mathrsfs}
\bibpunct{(}{)}{;}{a}{}{,}

\usepackage{siunitx}
\usepackage{dcolumn}
\newcolumntype{d}[1]{D{.}{.}{#1}}
\usepackage{verbatim}
\usepackage{float}
\def\mso{\,M_\odot}

 \def\mso{\,\mathrm{M}_\odot}

 \def\simle{\mathrel{\hbox{\rlap{\hbox{\lower4pt\hbox{$\sim$}}}\hbox{$<$}}}}
 \def\simgr{\mathrel{\hbox{\rlap{\hbox{\lower4pt\hbox{$\sim$}}}\hbox{$>$}}}}

\newcommand*{\koushik}{\textcolor[rgb]{0.4,0.7,0}} 

\begin{document}

\title{Reverse Algols and hydrogen-rich Wolf-Rayet stars from very massive binaries 
}


\author{K. Sen\inst{1,2}
\and N. Langer\inst{1,3}
\and D. Pauli\inst{4}
\and G. Gr\"afener\inst{1}
\and A. Schootemeijer\inst{1}
\and H. Sana\inst{5}
\and T. Shenar\inst{6,5}
\and L. Mahy\inst{7}
\and C. Wang\inst{8}
}

\institute{Argelander-Institut f\"ur Astronomie, Universit\"at 
Bonn, Auf dem H\"ugel 71, 53121 Bonn, Germany \\
\email{ksen@astro.uni-bonn.de, senkoushik1995@gmail.com}
\and Institute of Astronomy, Faculty of Physics, Astronomy and Informatics, Nicolaus Copernicus University, Grudziadzka 5, PL-87-100 Torun, Poland
\and Max-Planck-Institut f\"ur Radioastronomie, Auf dem H\"ugel 
69, 53121 Bonn, Germany
\and Institut f\"ur Physik und Astronomie, Universit\"at Potsdam, Karl-Liebknecht-Str. 24/25, 14476 Potsdam, Germany
\and Institute of Astronomy, KU Leuven, Celestijnenlaan 200D, 3001 Leuven, Belgium
\and Anton Pannekoek Institute for Astronomy, Science Park 904, 1098 XH, Amsterdam, The Netherlands
\and Royal Observatory of Belgium, Avenue circulaire/Ringlaan 3, B-1180 Brussels, Belgium
\and Max Planck Institute for Astrophysics, Karl-Schwarzschild-Strasse 1, 85748 Garching, Germany
}

\date{Received \today / Accepted ...}

\abstract
{Massive star feedback affects the evolution of galaxies, where the most massive stars
may have the largest impact. The majority of massive stars are born as members of close binary systems. Here, we investigate detailed evolutionary models of very massive binaries (30$\dots$90$\mso$) with Large Magellanic Cloud (LMC)  metallicity. We identify four effects defying the conventional knowledge of binary evolution, which are all related to the proximity of the models to the Eddington limit. We find that the majority of systems undergo mass transfer during core hydrogen burning. During the ensuing nuclear timescale evolution, many mass donors remain more massive than their companions (``reverse Algols''), and nuclear timescale mass transfer may be interrupted or absent all together. Furthermore, due to the elevated luminosity-to-mass ratio, many of the core-hydrogen burning donors may develop Wolf-Rayet type winds, at luminosities where single stars would not. We identify observational counterparts of very massive reverse Algol binaries in the LMC, and discuss their contribution to the observed hydrogen-rich Wolf-Rayet stars. We argue that an understanding of very massive Algol systems
is key to predict the advanced evolution of very massive binaries, including their ability to evolve into observable gravitational wave sources.

}

\keywords{stars: massive -- stars: evolution -- binaries: close -- Stars: Wolf-Rayet}

\maketitle

\section{Introduction}
\label{section_introduction}

Massive stars are known to affect multiple aspects of the evolution of our 
Universe \citep{haiman1997,maclow2004,langer2012}. They regulate star formation 
in galaxies \citep{Maclow2005,crowther2019} and drive chemical evolution
\citep{Burbidge1957,Pignatari2010,Thielemann2011,Kasen2017,Kajino2019}, 
where the dominant contribution may come from the most massive stars, particularly 
at sub-solar metallicity \citep{Kozyreva2014,crowther2016}. They further produce 
supernova explosions \citep{Burrows1995,Burrows2021,David2022b}, neutron stars 
\citep{Baym2018,Isaac2018}, stellar-mass black holes \citep[BHs,][]{Orosz2011,Miller-Jones2021} 
and gravitational waves events \citep{abbott2016,abbott2019,Abbott2021}. 

Recent observations have provided empirical evidence for massive 
stars to be preferentially born in binaries and higher-order 
multiples with at least one nearby companion 
\citep{sana2012,sana2013,sana2014,moe2017,Banyard2021}. The 
proximity of such companion has a strong impact on 
the evolution of both stars 
\citep{podsiadlowski1992,selma2013,DeMarco2017,kruckow2018,wang2020,Klencki2020}. 
In this situation, comparing binary evolution model predictions to 
observed populations of massive binaries \citep{Vanbeveren1998b,Han2003,selma2007, Eldridge2017,Menon2021,Abdul-Masih2021,Sen2021b,El-Badry2022,Patrick2022} 
is of paramount importance to place constraints on the uncertainties 
associated with massive star evolution \citep{langer2012,crowther2019}. 

The growing stellar radius during core hydrogen burning leads to
mass transfer via Roche-lobe overflow in short-period massive binaries  \citep{pols1994,vanbeveren1998,nelson2001,selma2007,Sen2021b}. 
This so-called Case\,A mass transfer is unique in the sense that it 
contains a nuclear timescale mass transfer phase (slow Case\,A), and 
has the Algol binaries \citep{surkova2004,malkov2020,Li2022} as 
long-lived observational counterparts. Slow Case\,A mass transfer occurs because 
after the mass ratio inversion during the foregoing fast (thermal timescale) 
Case\,A mass transfer phase, any further mass transfer widens the orbit \citep{Soberman1997,wellstein2001}. 
Consequently, after the mass donor becomes thermally relaxed, mass 
transfer is only driven by its expansion on the nuclear timescale. 
During this phase, the binary systems fulfill the so-called Algol 
paradox (named after $\beta$\,Per =  Algol, \citealp{Eggen1957}), 
with a less massive mass donor being more evolved than its more massive 
companion \citep{Paczynski1971,Batten1989,Pustylnik1998}.

Algols binaries present a distinct opportunity to test stellar and 
binary evolution physics, and in particular to constrain the mass transfer 
efficiency of the important fast mass transfer phase, during which the 
majority of the donor's envelope mass is removed \citep{selma2007,Sen2021b}. 
These systems are also expected to be the progenitors of short-period 
Wolf-Rayet (WR)+OB star binaries \citep{Hucht2001,petrovic2005}, high-mass 
X-ray binaries \citep{Valsecchi2010a,qin2019,quast2019,Sen2021a},
and, possibly, double compact binaries and merging black holes 
\citep{Bond1984,Voss2003,Belczynski2008,Dominik2012,Stevenson2017,Mandel2018, Chruslinska2018,kruckow2018,Spera2019,vigna-gomez2019,Antonini2020, Mapelli2020,Belczynski2020,Kremer2020,Monica2021,Marchant2021,Bavera2021, Broekgaarden2022,vanSon2022,Briel2022}. 
Only the detailed reproduction of the observed properties of massive binaries 
in the early stages of binary evolution \citep{Sen2021b,wang2020,Wang2022} 
can ensure that errors in the models for these stages do not propagate into 
our predictions of later binary evolution stages \citep[see, for e.g.][]{Olejak2021,Belczynski2022}. 

For initial primary star masses below $\sim$40$\mso$, detailed analyses of 
binary models following Case\,A evolution have been performed (see references above).
They essentially confirmed the classical picture of Algol binary evolution
derived from intermediate mass models \citep[e.g.][]{VanRensbergen2021}\footnote{We 
note that at low mass, when stars have convective envelopes, the character 
of Case\,A evolution is very different \citep{Giuricin1983,Lanza1999,Richards1993,Zavala2002}.}. 
While these models still cannot explain all of the individual massive 
Algol binaries \citep{selma2007,Sen2021b}, they are able to reproduce the 
overall properties of the majority of them, such as the distribution 
of orbital periods and mass ratios, implying that the evolutionary 
state of the observed Algol binaries is well understood. 

In contrast, detailed models of very massive binaries (with initial donor masses 
$\geq$\,40$\mso$) are sparse in the literature \citep{wellstein1999,petrovic2005,
Eldridge2017,shenar2020,Fragos2022,Pauli2022}, 
while rapid binary evolution models are generally unable to 
make accurate predictions for the Algol phase. Here, we try 
to remedy this by analysing models from a recent large grid of detailed models for 
very massive binary stars at LMC metallicity \citep{Pauli2022}, with focus 
on the evolution phase after the fast Case\,A mass transfer during which both components are still burning hydrogen in their cores. 

Our paper is organised as follows.
Section\,\ref{section_method} describes the physics assumptions used in 
the binary evolution models. In Sect.\,\ref{section_basic}, we discuss the 
salient features of Case\,A evolution in very massive binaries, and their 
relation to the Eddington limit. We estimate the initial binary parameter space for 
reverse Algol evolution in Sect.\,\ref{section_Pspace}, and show an animated 
view of reverse Algol evolution on the Hertzsprung-Russell (HR) diagram 
in Sect.\,\ref{section_animation}. We compare the models 
with observations of very massive semi-detached binaries and hydrogen-rich 
luminous WR stars in Sect.\,\ref{section_observations}. Finally, we discuss 
relevant uncertainties in massive binary modelling and their effects 
on our results in Sect.\,\ref{section_discussion}, before we summarise our 
work in Sect.\,\ref{section_conclusion}. 

\section{Method}
\label{section_method}

In this section, we briefly summarize the most relevant physics assumptions 
in the analysed detailed binary evolution models of \citet{Pauli2022}, while 
we refer to their work for more details. Furthermore, we describe here our 
method for assigning a optical depth parameter to a given stellar model, 
which will allow us to argue which of the models may correspond to WR type stars 
rather than to O\,stars. 

\subsection{Stellar physics}
\label{subsection_stellar_physics}

The models discussed in our paper have been calculated with version 10398 
of the one-dimensional stellar evolution code MESA \citep{mesa11,mesa13,mesa15,mesa18,mesa19}. 
The metallicity and initial chemical composition of the binary components 
correspond to that of the stars observed in young star-forming regions of 
the Large Magellanic Cloud (LMC). They are set as in \citet{brott2011}, 
with hydrogen, helium, and metal mass fractions of 0.7391, 0.2562 and 
0.0047, respectively, and account for the non-Solar metal abundance ratios 
in the LMC. Custom made OPAL opacity tables \citep{iglesias1996} for 
these initial abundances have been used. The models take into account differential 
rotation \citep{heger2000,Spruit2002}, magnetic angular momentum transport 
\citep{Spruit2002,heger2005} and rotational mixing via the Eddington-Sweet 
circulation, the Goldreich-Schubert-Fricke instability, and the secular and 
dynamic shear instability \citep{heger2000}. 

The adopted wind mass-loss rates depend on the surface hydrogen mass fraction 
$X_{\rm H}$ and effective temperature $T_{\rm eff}$ of the stellar model. 
For hot stars ($T_{\rm eff} > 25$\,kK) with $X_{\rm H} > 0.7$, the 
prescription of \citet{Vink2000} is used. For cooler stars ($T_{\rm eff} < 25$\,kK) 
with $X_{\rm H} > 0.7$, the maximum of the mass-loss rate of \citet{Vink2000} 
and \citet{Nieuwenhuijzen1990} is adopted. For stars with $X_{\rm H} < 0.4$, 
the \citet{Nugis2000} mass-loss rate is implemented with a wind clumping
factor D = 3 instead of 10 (\citealp{Pauli2022}, see also \citealp{yoon2017b}). 
Finally, for stars with $X_{\rm H}$ between 0.4 to 0.7, the mass-loss rate is
linearly interpolated between the \citet{Vink2000} rate and the reduced 
\citet{Nugis2000} rate. Mass loss rates for more evolved types of stars, 
such as hydrogen-free WR stars, are not relevant here. 

Regions of convective instability inside the star are determined using the 
Ledoux criterion. Convection is modelled using the standard Mixing Length 
Theory of \citet[MLT,][]{bohm1960}, with a mixing length parameter of 
$\alpha_{\rm MLT}$\,=\,1.5. In superadiabatic regions with a stabilizing molecular 
gradient, semi-convection is assumed to occur \citep{Langer1983,langer1991} 
with an efficiency parameter of $\alpha_{\rm sc}$ = 1 \citep{schootemeijer2019}. 
Thermohaline mixing is modelled following the work of \citet{cantiello2010}. 
Overshooting above the convective core is implemented as a step function 
extending the core by 0.335 times the local pressure scale height at the core boundary 
\citep{brott2011}. To account for the composition gradients in the 
rejuvenation process of mass gaining stars \citep{braun1995}, we only 
allow overshooting in regions with nearly constant composition \citep{pablothesis}. 

\subsection{Binary physics}
\label{subsection_binary_physics}

The binary models are calculated from the start of their hydrogen burning, 
assuming that both stars start burning hydrogen at the same time. The orbits 
are assumed to be circular, and the initial rotation period of the 
stars is set equal to the initial orbital period of the binary, with 
the spin axes perpendicular to the plane of the orbit. Time-dependent 
tides are modelled as in \citet{detmers2008}. The tidal synchronisation 
timescale is taken from the dynamical tide model of \citet{zahn1977}, since 
our work focuses on main sequence stars with radiative envelopes.

In the case of Roche-lobe overflow of one star, the mass transfer rate 
is calculated such that the donor star marginally fills its Roche lobe. 
When both stars fill their Roche lobes, we apply the contact binary scheme
from \citet{pablo2016}, also described in detail in \citet{Menon2021}. 
The evolution of models during the contact phase is stopped if the binary 
undergoes L2 overflow. Angular momentum accretion of the mass gaining 
star is based on the results of \citet{lubow1975} and \citet{ulrich1976}, 
implemented as in \citet{selma2013}, where a distinction is made between 
ballistic and disc modes of accretion. 

We assume that mass transfer in our binary models is conservative unless 
the mass accreting star spins up to critical rotation. When the accretor 
is critically rotating, we remove the excess transferred mass from the 
binary via an enhanced stellar wind, with a specific angular momentum 
equal to the specific orbital 
angular momentum of the accretor. Since in close binaries, tides impair  
the accretor's spin-up, the mass transfer efficiency in the models 
depends on the orbital period, with 
higher mass transfer efficiencies obtained for shorter-period binaries. 
Finally, when the energy required to remove the excess mass exceeds 
the combined luminosity of both stars, the model calculation is stopped.
For a comprehensive discussion of these physics assumptions, 
see \citet[][]{pablothesis} and \citet{Sen2021b}. 

\subsection{Optical depth parameter}
\label{subsection_optical_depth_method}

In order to derive the frequency-dependant emergent photon spectrum 
of a stellar model, detailed model atmosphere calculations are required 
\citep[e.g.][]{Groh2013,Jung2022}. While this is beyond the scope of 
our paper, we aim to assess which of the analysed stellar model would 
produce an emission-line-dominated spectrum and thus correspond to the 
class of Wolf-Rayet stars. To this end, we follow \citet{David2022a}
and \citet{Pauli2022} to compute the optical depth parameter of the 
adopted stellar winds. 

We follow \citet{DeLoore1982} and \citet{Langer1989} in assuming a 
$\beta$=1 wind-velocity law \citep{Vink2000} to estimate the optical 
depth of a stellar wind with mass loss rate $\dot{M}$ for a star with 
radius $R$ as
\begin{equation}
    \tau_{\rm wind}(R) = \frac{\kappa |\dot{M}|}{4\pi R (\upsilon_{\infty} - \upsilon_{0})} \mathrm{ln}\left( \frac{\upsilon_{\infty}}{\upsilon_{0}}\right) ,
    \label{eqn:tau}
\end{equation}
using the electron scattering opacity as $\kappa = 0.2 (1+X_{\rm H})$\,cm$^{2}$\,g$^{-1}$. 
Here, $\upsilon_{\infty}$ is the terminal wind speed, $\upsilon_{0}$ 
is the expansion velocity near the stellar surface taken as 20\,km\,s$^{-1}$, 
and $X_{\rm H}$ is the mass fraction of hydrogen at the stellar surface. 
For stars with $4.4 < \log\,T_{\rm eff} < 4.7$, we assume 
$\upsilon_{\infty} = 2.6 \upsilon_{\rm esc}$, and for cooler stars
$\upsilon_{\infty} = 1.3 \upsilon_{\rm esc}$ \citep[as in][]{Pauli2022}. 
For the escape velocity $\upsilon_{\rm esc}$, we account for the electron 
scattering Eddington factor given by 
\begin{equation}
 \Gamma_{\rm e}=\frac{\kappa L}{4\pi c GM}
  = 10^{-4.813} \times \left( 1 + X \right) \frac{L/L_\odot}{M/M_\odot},
\label{eq:GAMMA_E}
\end{equation}
where $L$ and $M$ are the luminosity and mass of the star, 
respectively. We ignore the dependence of the terminal wind 
speed on metallicity $Z$ ($\upsilon_{\infty} \propto Z^{0.1-0.2}$, 
\citealp{Vink2021,Marcolino2022}), as it is not expected to 
affect our results significantly. 

Due to the rough approximations required for our approach (e.g. on the 
opacity), we do not expect the optical depth parameters 
computed for the models to accurately
represent the true optical depth in the corresponding stellar 
winds. However, they include the main dependencies on 
mass loss rate, wind velocity and stellar radius, and as such
they may be valid order-of-magnitude estimates.
Moreover, these parameters can be meaningfully compared relative 
to each other, that is, for different stellar models, and, importantly,
to the optical depth parameters computed in the same way for 
observed stars. 
From this ansatz, \citet{David2022a} and \citet{Pauli2022} find
a threshold optical depth parameter for hydrogen-free Wolf-Rayet stars
of $\tau_{\rm wind} \simeq 1.5$, and \citet{Pauli2022} suggest 
a significantly smaller value for hydrogen-rich Wolf-Rayet stars.
We show in Sect.\,\ref{obs2} that for the models
investigated here, a threshold value as low as $\tau_{\rm wind} \simeq 0.1$ 
may be appropriate.

\section{Distinct effects in the evolution of very massive Case\,A binaries}
\label{section_basic}

In this section, we highlight four fundamental effects which occur only 
in the evolution of very massive binaries. The mass limits above which 
they exhibit these effects is gauged here based on the LMC binary 
model grid, but they may 
depend on metallicity. All four effects are related to the proximity of 
very massive stars to their Eddington limit, and therefore occur 
naturally in any detailed binary evolution models. The reverse Algol 
configuration in particular can be found in models of \citet{Sybesma1986} 
and of \citet{Stanway2018}, but the unique implications of this 
evolutionary path we raise here were not identified before. 
We describe specific binary models in detail in the appendix 
to illustrate our findings and only discuss their underlying physics here. 
In Sect.\,\ref{section_Pspace}, we identify the initial binary parameter 
space in which the reverse Algol scenario (Sect.\,\ref{section_r}) is 
expected to play a role. We discuss uncertainties in our inherent assumptions 
in Sect.\,\ref{section_discussion}. 

\subsection{Most very massive binaries undergo Case\,A mass transfer}
\label{section_A}

Massive main sequence stars ($\sim$20\,$\mso$) typically expand by a factor 
of three during their main sequence evolution. Case\,A mass transfer is 
therefore usually associated with orbital periods of the order of a few 
days (Fig.\,5 of \citealp{Sen2021b}). But, due to the proximity to 
the Eddington limit and corresponding envelope inflation in very massive 
core-hydrogen burning stars \citep{brott2011,Grafener2012,sanyal2015,sanyal2017}, 
the limiting orbital period for Case\,A evolution increases sharply at higher 
mass. For example, \citet{Pauli2022} find this limiting initial orbital period 
in LMC binary models to be at 16\,d, 120\,d and 2000\,d for initial donor 
masses of $32\mso$, $50\mso$ and $56\mso$, respectively. This can be 
understood by considering single-star evolution models. In the LMC 
models of \citet{brott2011}, the initial mass at which envelope inflation 
shifts the terminal age main sequence (TAMS) effective temperature to about 
halfway between the zero-age main sequence (ZAMS) and the Hayashi line, 
is about $40\mso$. Consequently, at higher mass, the 
majority of all interacting binaries will undergo Case\,A mass transfer. 

Due to the large luminosity-to-mass ratio of very massive stars, it is 
also more likely that very massive binaries can undergo highly non-conservative 
mass transfer without merging. In the models studied here, it is assumed 
that a merger occurs when the available photon energy is insufficient to 
push the transferred mass that can not be accreted by the donor to infinity 
\citep{pablothesis}. As this condition is more easily avoided for more massive 
stars, we find that most of our very massive Case\,A binaries with initial 
periods even up to $\sim 1000\,$d can avoid merging. In the appendix, we show 
a detailed example for a very massive binary with an initial orbital period of 
100\,d (Fig.\,\ref{fig:sd_reverse}).

\subsection{Donors may remain the more massive binary component}
\label{section_r}

In the classical picture of Case\,A binary evolution in intermediate and 
massive stars, once mass transfer commences it occurs on the rapid, thermal 
timescale. The reason is that any transfer of mass from the donor to the 
accretor leads to a shrinking orbit until the mass ratio is inverted. Since 
for the mass donor to remain within its Roche volume, it needs to be more 
compact than it can be in thermal equilibrium, fast mass transfer can only 
end after the mass ratio has inverted and the orbit widens again.

Also in very massive binaries, Case\,A mass transfer starts on the thermal 
timescale. However, in those, the fast mass transfer phase can end before 
mass ratio inversion, for two reasons. Firstly, fast Case\,A mass transfer 
ends when the donor is stripped so far that helium-enriched matter appears 
at its surface, since from that moment on any mass loss from the donor leads 
to a decreasing thermal equilibrium stellar radius. And secondly, as 
discussed in the next subsection, the donor's stellar wind mass loss can become 
so strong that the orbit widens already before the mass ratio is inverted. 
We discuss consequences for the slow (nuclear timescale) Case\,A
mass transfer in Sect.\,\ref{section_d}.

Surface helium enrichment before mass ratio inversion in very massive binaries
occurs due to the large convective core fractions of very massive stars 
--- an effect of their large Eddington factors. The fraction of the 
total mass which forms the unprocessed envelope of a massive star is decreasing 
with mass. The convective core mass fractions (excluding the overshooting 
region) at the beginning of core hydrogen burning are roughly 0.3, 0.5, and 
0.8 for stars of $10\mso$, $30\mso$ and $80\mso$, respectively. Therefore, 
in binaries with $10\mso$ donors, the mass ratio of the binary will invert 
long before helium-enriched matter appears at its surface. In contrast, for 
stars above $\sim 30\mso$, mass transfer via Roche-lobe overflow may remove 
the envelope of the donor without producing a mass ratio inversion of the 
binary. The consequence is a nuclear timescale Case\,A mass transfer phase, 
during which the more massive star is transferring mass to the less massive 
companion, that is, a reverse Algol configuration (see Fig.\,\ref{fig:sd_reverse}). 

Reverse Algol evolution becomes more likely the higher the initial mass of the 
donor star. While this is so due to the larger convective core mass fraction 
(see above), another effect corroborates this. We find that mass transfer
may be less conservative in more massive binaries, since more energy is available 
to remove mass from the binary system. This allows
binaries with more extreme initial mass ratios to avoid merging.  
Results from the recent detailed binary model grid of 
\citet{Pauli2022} show that binaries with initial 
companion masses of 50\%, 30\%, and 20\% of the initial donor mass can still 
survive Case\,A mass transfer for initial donor masses of $32\mso$, $50\mso$, 
and $80\mso$, respectively. We explore the Reverse Algol parameter space more
comprehensively in Sect.\,\ref{section_Pspace}.

While also the ordinary Algol evolution occurs in the analysed models,
i.e., they do undergo a mass ratio inversion during fast Case\,A mass transfer,
in some of them the donor remains the more luminous stars
(Appendix\,\ref{appendix:more_examples}). We find models of such binaries 
in both, the semi-detached and the detached configuration 
(Figs.\,\ref{snapshot} and\,\ref{fig:de_normal}). The less massive yet more luminous star 
may also show WR characteristics (Sect.\,\ref{section_wr}), as we 
expect its optical depth parameter may be similar to that of observed 
hydrogen-rich WN stars (see the wind mass-loss rates in 
Fig.\,\ref{fig:de_reverse} and Fig.\,\ref{fig:de_normal}). 

\begin{figure*}
    \centering
    \includegraphics[width=\hsize]{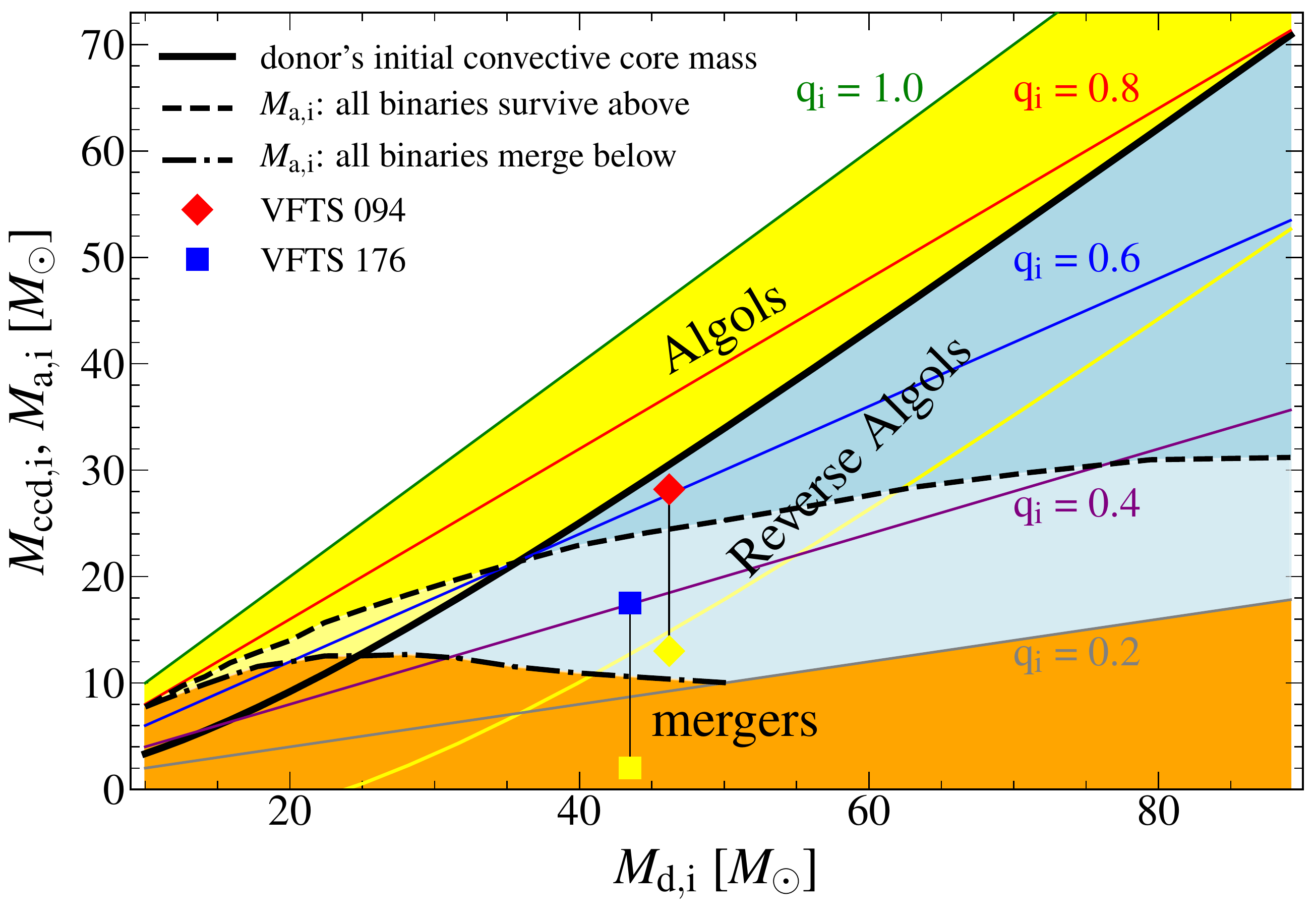}
    \caption{Initial convective core mass of the mass donor ($M_{\rm ccd,i}$ thick black line) and initial accretor mass $M_{\rm a,i}$ (thin colored lines) for five initial mass ratios $q_{\rm i} = M_{\rm a,i}/M_{\rm d,i}$ (0.2, 0.4, 0.6, 0.8, 1.0), as a function of the initial donor mass $M_{\rm d,i}$ in massive binary systems. 
    For highly non-conservative mass transfer, reverse Algol evolution can not occur above the thick black line (yellow region) but below (blue region), whereas for conservative evolution it can not occur above the yellow line. 
    The symbols mark the positions of VFTS 094\,and VFTS\,176 in this diagram, assuming inefficient (blue) or conservative (yellow) mass transfer, with the ordinate value giving their initial accretor masses.
    The black dashed line shows the limiting initial companion mass above which all models of \citet{Pauli2022} which avoid contact also avoid merging during Case\,A mass transfer, and the dash-dotted line shows the limiting initial companion mass below which all models merge. Between these two lines, 
    whether the models merge or not is a function of their initial orbital period, and
    the fraction of models that merge increases for lower initial accretor masses. 
    Lighter shading is used to indicate that not all models avoid merging. It is also assumed that all binaries with q$_{\rm i} < $ 0.2 (grey line) merge before entering the Algol stage.   
     }
    \label{fig:inversion_pspace}
\end{figure*}

\begin{figure}
    \centering
    \includegraphics[width=\hsize]{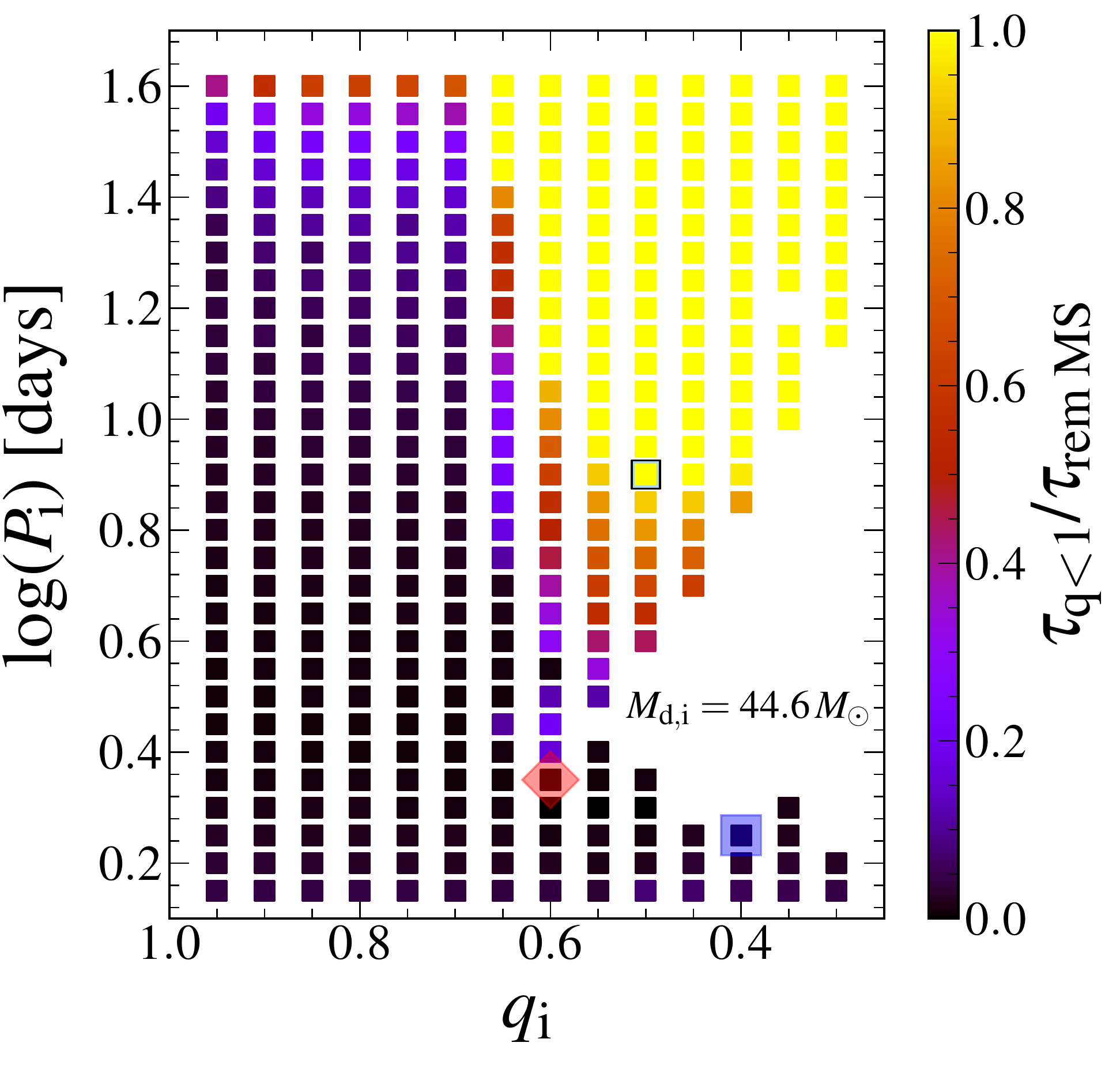}
    \caption{Initial orbital period and initial mass ratio of the models of the considered binary evolution grid with an initial donor mass of 44.6\,$\mso$ (small squares). 
    We only show models that do not merge during the fast Case\,A mass transfer \citep[see Figs.\,1 and\,F1 of][]{Sen2021b}. The colour indicates the fraction of the lifetime of donor after the fast Case\,A mass transfer phase during which the donor is more massive than the accretor (see color bar to the right). The model discussed in the right panel of Fig.\,\ref{fig:de_reverse} is highlighted with a black frame. The large red diamond and large blue square show the estimated positions of VFTS\,094 and VFTS\,176 {assuming highly 
    non-conservative mass transfer}, respectively (see text).  }
    \label{fig:inversion_models}
\end{figure}

\subsection{Donors may obtain WR type winds}
\label{section_wr}
As the wind mass-loss rate for O\,type stars increases steeply with luminosity 
\citep[see, for e.g., Eq.\,12 of][]{Vink2000}, and the optical depth parameter 
is proportional to the wind mass-loss rate (Eq.\,\ref{eqn:tau}), one may expect 
that moving up in mass, eventually even unenriched main sequence stars have 
winds with WR characteristics. Indeed, luminous and very hydrogen-rich WR 
stars --- with surface hydrogen mass fractions of up to $X_{\rm H}=0.7$ --- 
are observed and often interpreted as core hydrogen burning single stars 
\citep{Hainich2014}. We designate the stellar mass above which unenriched 
main-sequence single stars have WR-type winds (here at LMC metallicity) as $M_{\rm WNH}$. 

In very massive binaries, as we discussed above, the fast mass transfer reduces 
the stellar mass to the initial convective core mass of the mass donor, such that 
when the fast mass transfer ends, the donor's hydrogen surface abundance is still 
close to its initial value (Fig.\,\ref{fig:sd_reverse}). At this stage, the donor's 
luminosity is close to its luminosity before the fast mass transfer 
(Fig.\,\ref{fig:hrds} in Sect.\,\ref{subsection_HRDs}). This is linked to the 
mass-luminosity relation for
core hydrogen burning stars, where a smaller mass implies a smaller luminosity,
but an increased average mean molecular weight implies a larger luminosity, such 
that both dependencies counteract each other (Fig.\,17 in 
\citealp{Kohler2015}). 
Consequently, our donor stars have an elevated luminosity-to-mass ratio, that is, 
they are closer to their Eddington limit compared to single stars of the same 
mass. This implies that our yet unenriched donors may develop WR-type 
winds at masses well below the corresponding single star mass limit $M_{\rm WNH}$.  

The very massive Case\,A donors evolve after the fast mass transfer 
with a surface hydrogen mass fraction close to their initial surface hydrogen mass fraction. Further-on during 
core hydrogen burning, the slow mass transfer and/or the 
enhanced stellar wind lead to decrease of their surface hydrogen mass fraction, and to 
a corresponding increase in their surface helium mass fraction. 
We may therefore expect very massive Case\,A 
donors with surface hydrogen mass fraction $X_{\rm H}$ in the range $0.7\dots0.2$
(Fig.\,\ref{fig:sd_reverse}), with WR-type winds.

In Sect.\,\ref{section_r}, we have seen that the donors may remain the more massive 
components in the binaries. As on top of that, they also have the larger average 
mean molecular weight, they may be far more luminous than their companion, which 
therefore may be difficult to spot in the combined spectrum. Since some of our 
binary models with such parameters have long orbital periods, it could be 
difficult to detect the binary status in corresponding observed systems.

\subsection{Donors may underfill their Roche-lobes}
\label{section_d}

The classical Case\,A mass transfer evolution leads to donor stars which, after 
the fast Case\,A mass transfer, keep transferring mass on the nuclear timescale 
throughout the remainder of their core hydrogen burning evolution 
\citep{pols1994,wellstein2001,Sen2021b}. 
In the very massive Case\,A binaries, this may be different, due to 
the high stellar wind mass-loss of the donors \citep{petrovic2005}. 
We find that in many of the investigated models, the wind mass-loss rate
exceeds the nuclear timescale mass transfer rate (Fig.\,\ref{fig:sd_reverse}).

In this situation, the stellar mass loss may indeed slow down the nuclear timescale 
radius growth of the donor, or --- as the donor becomes ever more helium-enriched 
at the surface --- even reverse it (right panel of Fig.\,\ref{fig:sd_reverse}). 
On top of that, despite mass being transferred to the less massive star in the 
reverse Algol situation, the stellar wind mass loss may lead to a widening of 
the orbit and an increase of the orbital period (see Sect.\,\ref{subsection_period}). 
We indeed find that only $\sim$20\% of the donors in the very massive models fill 
their Roche lobes, and in some cases the Roche lobe filling factors drop 
considerably during the core hydrogen burning evolution. 

\section{The parameter space for Reverse Algol evolution}
\label{section_Pspace}

The effects discussed in Sect.\,\ref{section_basic} occur only in stellar models which are close to their 
Eddington limit. This translates naturally to a limiting donor mass, below which 
we expect these effects do not occur. However, this mass limit can be different 
for the four different mentioned effects, and some will also depend on the 
initial stellar metallicity and other physics assumptions. Consequently, we can 
not comprehensively derive the initial binary parameter space where the effects 
are operating. However, we can quantify the parameter space for reverse Algol evolution through our Fig.\,1.


 
Figure\,\ref{fig:inversion_pspace} shows the donors initial convective core mass
$M_{\rm ccd,i}$ as function of their initial mass $M_{\rm d,i}$, which is the key quantity to estimate
whether a reverse Algol phase follows the fast mass transfer in a given binary system.
For the case of highly inefficient mass transfer (none of the 
transferred matter remains on the accretor), models in which the initial 
convective core mass of the donor exceeds the initial mass of the companion 
may become reverse Algols, since the donor is stripped to its initial convective 
core mass during fast Case\,A mass transfer. The borderline for reverse Algol evolution
is thus defined as 
\begin{equation}
    M_{\rm a,i} = M_{\rm ccd,i} , 
\end{equation}
where $M_{\rm a,i}$ is the initial mass of the accretor, or
\begin{equation}
\label{a}
    q_{\rm i}=f_{\rm ccd,i} ,
\end{equation}
with $q_{\rm i}$=$M_{\rm a,i}/M_{\rm d,i}$ and $f_{\rm ccd,i}$=$M_{\rm ccd,i}/M_{\rm d,i}$. It is expressed by 
the thick black line in Fig.\,\ref{fig:inversion_pspace}. Only binaries above it can invert their 
mass ratio and develop an ordinary Algol phase, whereas models below that 
line may become reverse Algol systems. 

For the case of conservative mass transfer, the borderline between ordinary and reverse 
Algol evolution is defined by the accretor's mass after fast Case\,A mass transfer
equaling the donor's initial convective core mass, i.e.,
\begin{equation}
    M_{\rm a,i} + (M_{\rm d,i} - M_{\rm ccd,i}) = M_{\rm ccd,i} , 
\end{equation}   
which implies
\begin{equation}
\label{b}
    q_{\rm i}=2f_{\rm ccd,i} - 1 .
\end{equation}
This condition
corresponds to the yellow line in Fig.\,\ref{fig:inversion_pspace}.  

These two borderlines for reverse Algol evolution are model-independent. 
They are not affected by stellar wind mass loss of the mass donor as long as 
the wind does not uncover helium-enriched layers, but wind mass loss of the mass 
gainer before the onset of mass transfer might shift both lines slightly upwards. 
While the reverse Algol parameter space shrinks for more efficient mass transfer, 
even for conservative evolution we still expect reverse Algol systems for 
sufficiently high donor masses. The reason is that the initial envelope mass 
fraction of the donor becomes ever smaller for higher initial masses. Below, 
we discuss the reverse Algol parameter space possible for 
inefficient mass transfer in more detail, which corresponds closer to the investigated
grid of binary evolution models. 

The thin coloured lines in Fig.\,\ref{fig:inversion_pspace} indicate selected initial binary mass ratios  
as function of the initial donor mass. For a given initial mass ratio,
reverse Algols can form only to the right of the intersection of these 
lines with the line giving the initial convective 
core mass of the donor. We can find a line of 
constant initial mass ratio which intersects with the line for 
the donor's initial convective core mass for any chosen donor mass.
The corresponding initial mass ratio is the maximum value for which
reverse Algol evolution can occur at this donor mass. 
For example, we can see that binaries with an initial donor mass above $\sim$35\,$\mso$
are not expected to invert their mass ratio during fast Case\,A mass 
transfer if their initial mass ratios are below$\,\sim$0.6.

For non-conservative mass transfer, the yellow and blue areas in 
Fig.\,\ref{fig:inversion_pspace} indicate where ordinary or reverse Algol 
evolution is possible. However,
when an Algol or reverse Algol evolution is indicated for a given 
$(M_{\rm d,i},M_{\rm a,i})$ in Fig.\,\ref{fig:inversion_pspace}, some or even 
all such models may merge before the (reverse) Algol phase is reached, 
depending on their initial orbital period. The dashed and dashed-dotted lines 
in Fig.\,\ref{fig:inversion_pspace} indicate this for the detailed binary model 
grid of \citet{Pauli2022}. 
They imply that --- when ignoring the shortest period binaries which 
develop contact and merge (\citealp[see Fig.\,1 and Fig.\,F1 of][]{Sen2021b}, 
see also \citealp{Menon2021}) --- all models above the dashed line avoid merging, 
and all models below the black dash-dotted line merge. In between these two 
lines (indicated by lighter shading), the fraction of surviving models drops 
from one to zero, with shorter period Case\,A models surviving near the dashed 
line but merging near the dash-dotted line (Fig.\,\ref{fig:inversion_models}). 
The binary models with small initial mass ratios ($q_{\rm i} < 0.2$) and 
those with the shortest initial orbital periods ($P_{\rm i} < 2\dots 3\,$d) 
are assumed to merge during the fast Case\,A.


Figure\,\ref{fig:inversion_models} shows the fraction of the time span from 
the end of fast Case\,A mass transfer to core hydrogen exhaustion of the 
donor star during which its mass exceeds that of the accretor, for 
models with an initial donor mass of 44.6\,$\mso$. The 
black-dominated region corresponds to models with a long-lived ordinary Algol 
phase while other colours indicate a long-lived reverse Algol evolution.
For initial periods above $\sim 4\,$d, the border between both regions corresponds roughly to $q_{\rm i} = 0.65$.
When we compare this to Fig.\,\ref{fig:inversion_pspace}, 
we obtain a very similar answer. The initial convective core mass of 44.6\,$\mso$ 
stars is about $28.5\mso$, which corresponds to a critical initial donor 
mass for reverse Algol evolution of the same value, that is, to a critical initial 
mass ratio of 0.64. 

Figure\,\ref{fig:inversion_models} also shows the limitation of our simple 
approach. It reveals a slight dependence of the critical mass 
ratio on the initial orbital period for initial periods above $4\,$d,
and a shift of the critical mass ratio towards smaller values for shorter initial periods.
The main reason is that in the shorter 
period binaries, tides may delay or prevent the spin-up of the 
accretor, which increases the mass transfer efficiency \citep[see Fig.\,F2 of][]{Sen2021b}, such that the critical mass ratio shifts from
Eq.\,\ref{a} to Eq.\,\ref{b}.

For the models with an initial donor mass of $\sim$44.6\,$\mso$, about 40\% 
of the surviving Case\,A binary models undergo reverse Algol evolution. 
Therefore, we expect observable counterparts of such models, in which  
overluminous and near Roche-lobe filling primaries orbit less massive and 
less luminous secondary stars. 

\begin{figure*}
\centering
\includegraphics[width=0.48\hsize]{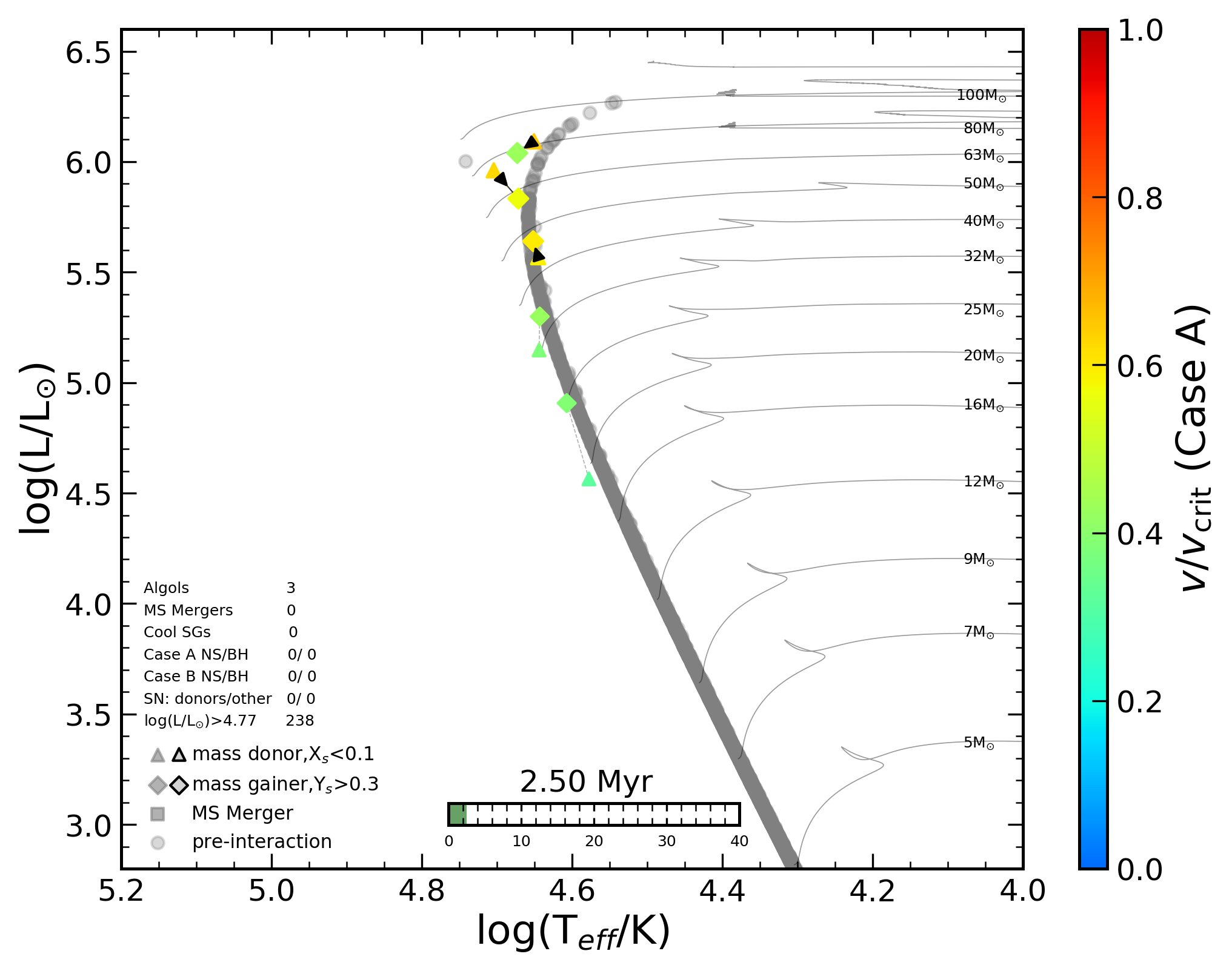}
\includegraphics[width=0.48\hsize]{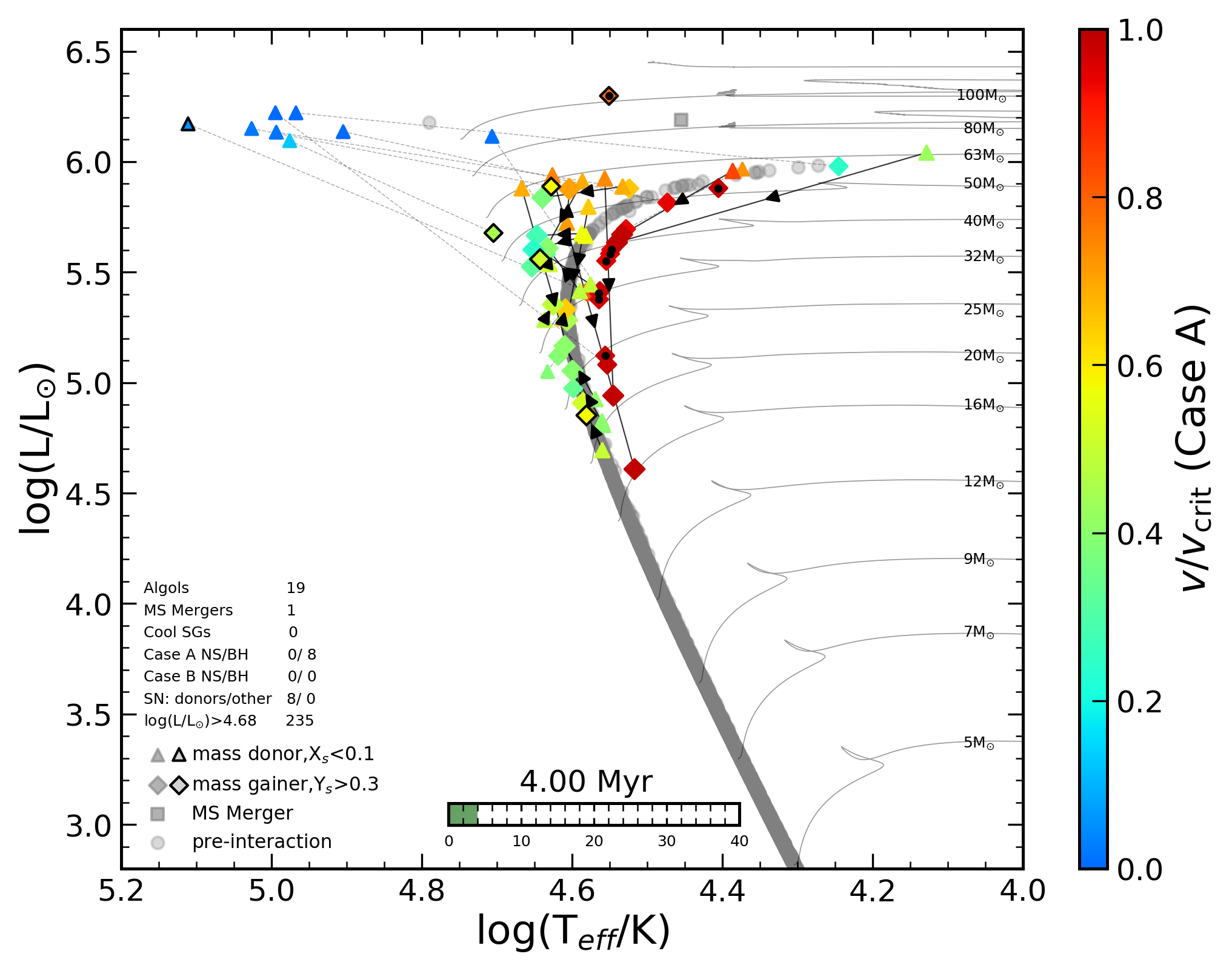}
\includegraphics[width=0.48\hsize]{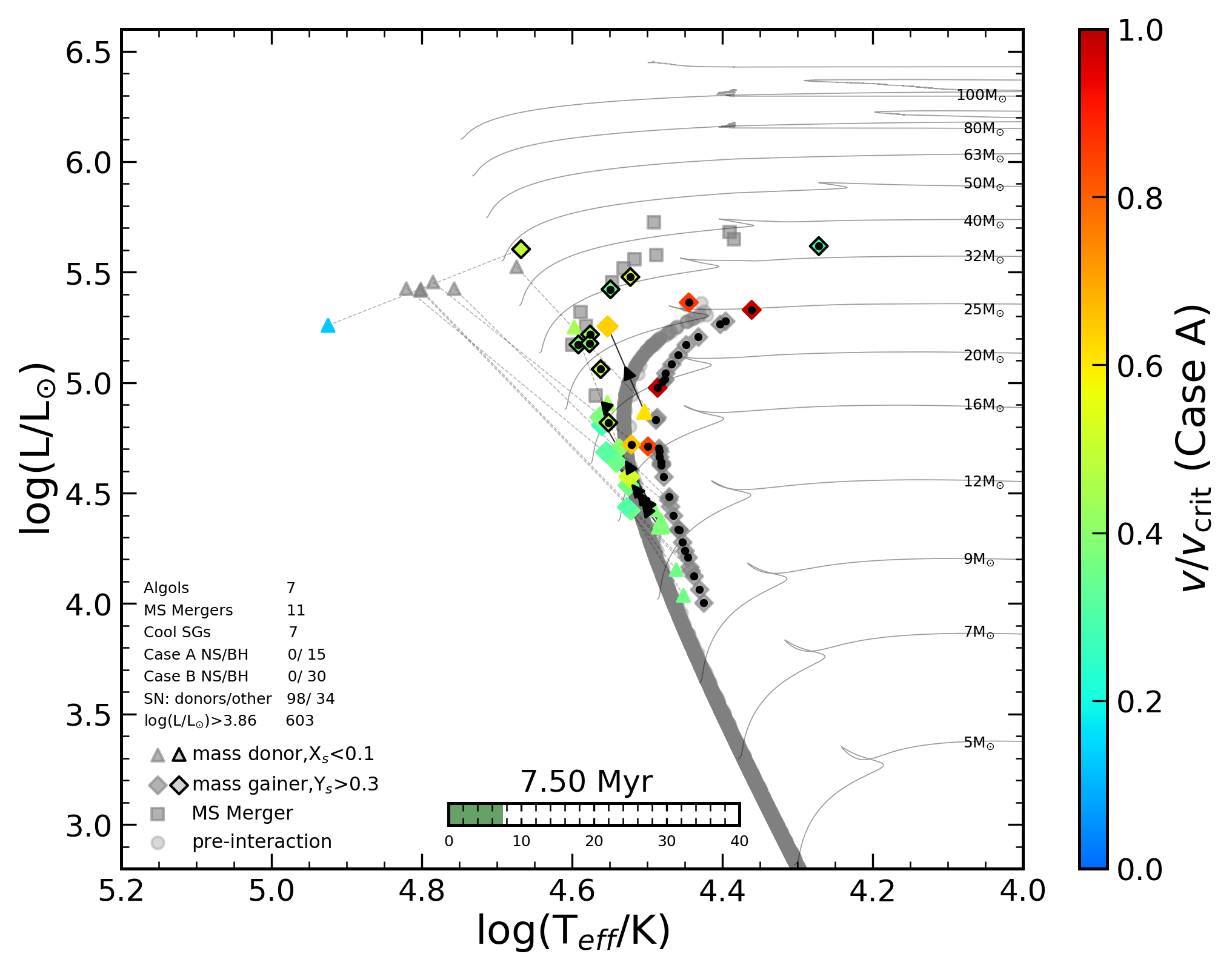}
\includegraphics[width=0.48\hsize]{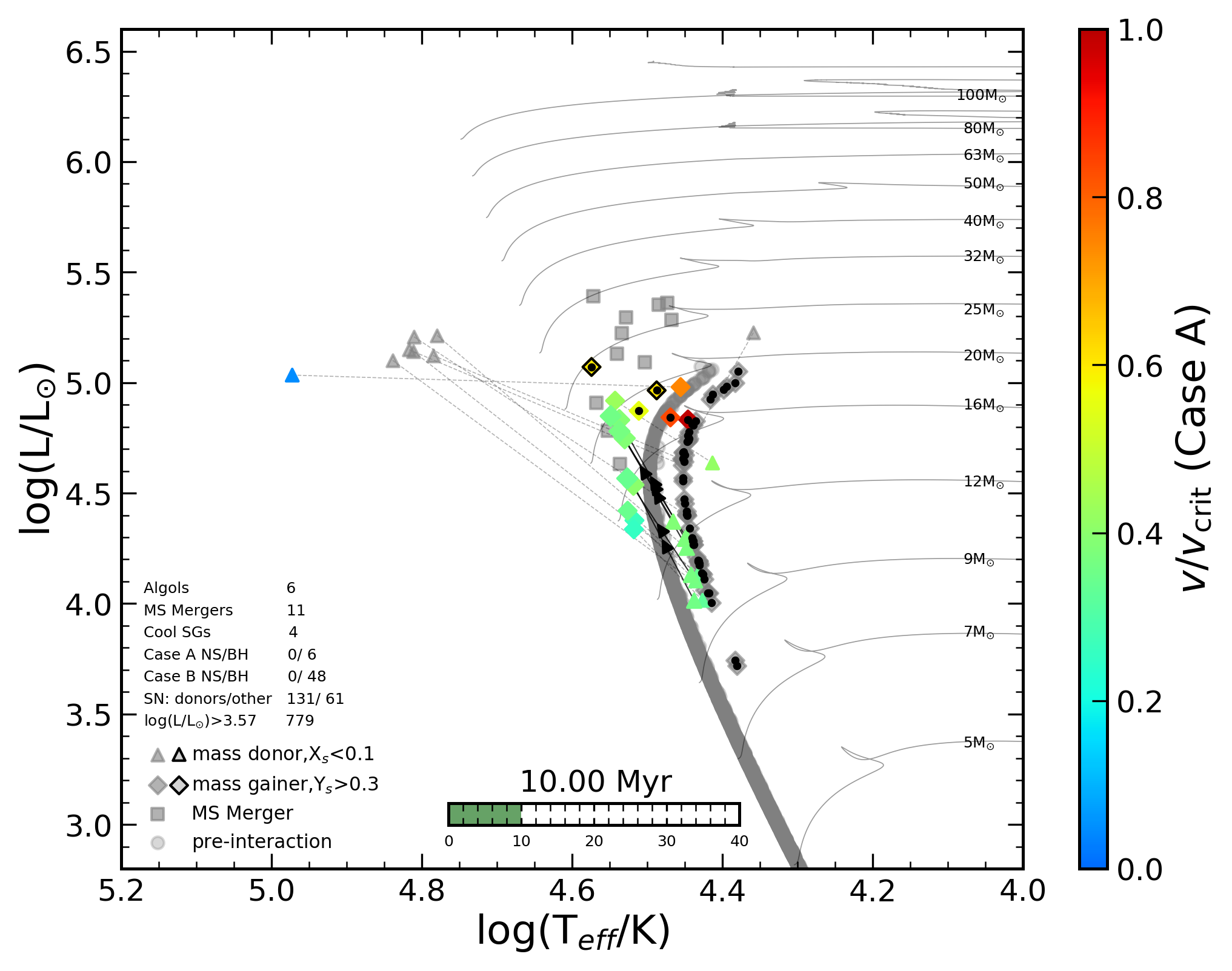}
\caption{Four snapshots of the animation showing the positions 
of both components of detailed massive binary models in the HR diagram, 
considering a coeval population following a Salpeter initial mass function 
and flat initial distributions of mass ratios and logarithms of the initial 
orbital periods. The snapshots correspond to an age of 2.5\,Myr (\textit{top 
left panel}), 4\,Myr (\textit{top right panel}), 7.5\,Myr 
(\textit{bottom left panel}) and 10\,Myr (\textit{bottom right panel}). 
Translucent grey circles indicate components of pre-interaction binaries, 
grey squares indicate Case\,A merger 
products, and mass donors and accretors after the onset of mass transfer are indicated 
by triangles and diamonds, respectively. 
Components of Case\,A binaries after the onset of mass transfer
are shown in colour, denoting the fraction of critical rotation 
rotation ($\mathrm{v}/\mathrm{v}_{\mathrm{crit}}$). 
The two components of semi-detached binaries 
are connected by solid black lines with an arrow indicating 
the direction of mass transfer. 
The two components of binaries that have interacted in the past 
are connected with grey dotted lines.
A black frame around a mass donor symbol (triangle) indicates a surface 
hydrogen mass fraction below 0.1, and a 
black frame around the symbol for a mass accretor (diamond) 
indicates a surface helium mass fraction above 0.3. 
The current age of the population is 
displayed in the center bottom with a time bar that 
fills up as the animation moves forward in time. The table 
above the legend indicates (from top) the current number of
i) Algol binaries, i.e., models in the slow Case\,A mass transfer phase; 
ii) core hydrogen burning main sequence merger products and, 
iii) cool red supergiants (log\,T$_{\mathrm{eff}}$\,<\,3.7); iv) and v)
Case\,A or Case B main sequence mass gainers, respectively, with a neutron star (white dot on symbol) or black 
hole (black dot on symbol) companion,
vi) core collapse events that occurred up to the current age, and
vii) pre-interaction binaries brighter than the indicated luminosity threshold,
which is 1.5\,dex below the luminosity of the brightest hydrogen-burning single star 
at the current age. }
\label{snapshot}
\end{figure*}

\section{An animated view of massive Algol models}
\label{section_animation}

Here, we discuss the evolution of massive Algol binaries in the 
Hertzsprung-Russell diagram through the animation 
of an coeval population 
of massive binary stars. We use the detailed binary evolution 
models introduced in \citet{wang2020}, which were calculated 
at a metallicity suitable for the Small Magellanic Cloud (SMC). 
In Sect.\,\ref{subsection_metallicity}, we discuss briefly the 
effect of metallicity on our results. 

The grid consists of 2078 binary evolution models with initial primary 
masses greater than $5\mso$, covering an initial mass ratio 
range of 0.3-0.95 and initial orbital periods of 1 day to 8.6 yrs. 
A Monte Carlo method was used 
to sample initial binary model parameters assuming a Salpeter 
initial mass function \citep[][]{salpeter1955}, a flat 
distribution of mass ratios and a flat distribution of the 
logarithm of initial orbital periods. The stellar physics 
assumptions are the same as in \citet{brott2011} and the 
binary physics assumptions are the same as in \citet{Sen2021b}. 
For more details, we refer to \citet{wang2020}. 

Our animation shows the positions of both binary components 
for the coeval population of binary stars in the HR diagram 
and covers their first 40\,Myr of evolution. An interpolation 
of stellar parameters between the binary evolution models is 
not needed. Only an interpolation in time for each binary model 
was performed. Due to the high time resolution of the MESA models, 
this did not lead to noticeable errors in the animation 
\citep{wang2020}. Figure\,\ref{snapshot} shows four 
snapshots of the animation to describe its features. 

Since thermal timescale mass transfer is too fast to be resolved 
in the animation, the highlighted mass-transferring binaries 
are binaries undergoing slow Case\,A mass transfer. The first 
such case appears with a donor mass near $40\mso$ at $t\sim 1.55\,$Myr, 
in the classical Algol configuration. Around $t\sim2.5\,$Myr (Fig.\,\ref{snapshot}, 
top left panel), we begin to see binary models in the 
semi-detached configuration where the more luminous star is 
transferring mass to a less luminous companion. 

With time, more semi-detached binaries appear where the more 
luminous donor is found to transfer mass to a less luminous 
companion. For example, at 4 Myr (Fig.\,\ref{snapshot}, 
top right panel), we find 19 binaries in the semi-detached 
configuration, of which 11 are in the classical Algol 
configuration and eight are in the reverse Algol configuration. 
We see that there are more semi-detached binaries in the 
reverse Algol configuration at higher luminosities than at 
lower luminosities. There are no reverse Algol binaries 
below log (L/L$_{\odot}$) = 5.3, because 
less luminous stars are generally less massive, and 
the parameter space for reverse Algol evolution decreases 
for lower masses (Fig.\,\ref{fig:inversion_pspace}). 

Our animation shows further that in ordinary as well as in 
reverse Algols, the donor star is generally cooler than the 
accretor. This implies that in ordinary Algols, the bolometric 
correction of the donor will be less than that of the more 
luminous accretor, which helps to identify the donor in the 
combined spectrum. In reverse Algols, the effect goes the other 
way. The less luminous accretor is also hotter and may therefore 
be hard to identify spectroscopically. For example, a donor 
star of a reverse Algol model in the right panel of Fig.\,\ref{snapshot} 
is found at $T_{\rm eff}\simeq 11\,$kK, while the accretor 
has $T_{\rm eff}\simeq 37\,$kK. Together with a luminosity 
ratio of 6.3, it implies that the mass donor is about 5.2\,mag 
brighter in the visual spectral range, and will therefore 
completely outshine the mass gainer.

As time progresses in our animation, the initial masses of the stars 
which evolve into semi-detached systems decrease, and the 
corresponding initial mass ranges move below the initial mass threshold 
for reverse Algol formation. We find that the last reverse Algol model 
disappears after an age of $\sim$7.5\,Myr (Fig.\,\ref{snapshot}, left 
bottom panel), whereafter all semi-detached binaries appear in the 
ordinary Algol configuration (Fig.\,\ref{snapshot}, right bottom 
panel). This time corresponds to the hydrogen-burning lifetime of a 
$25\mso$ star, and is thereby in good agreement with the simple 
prediction derived from Fig.\,\ref{fig:inversion_pspace}.



\begin{figure*}
    \centering
    \includegraphics[width=0.48\hsize]{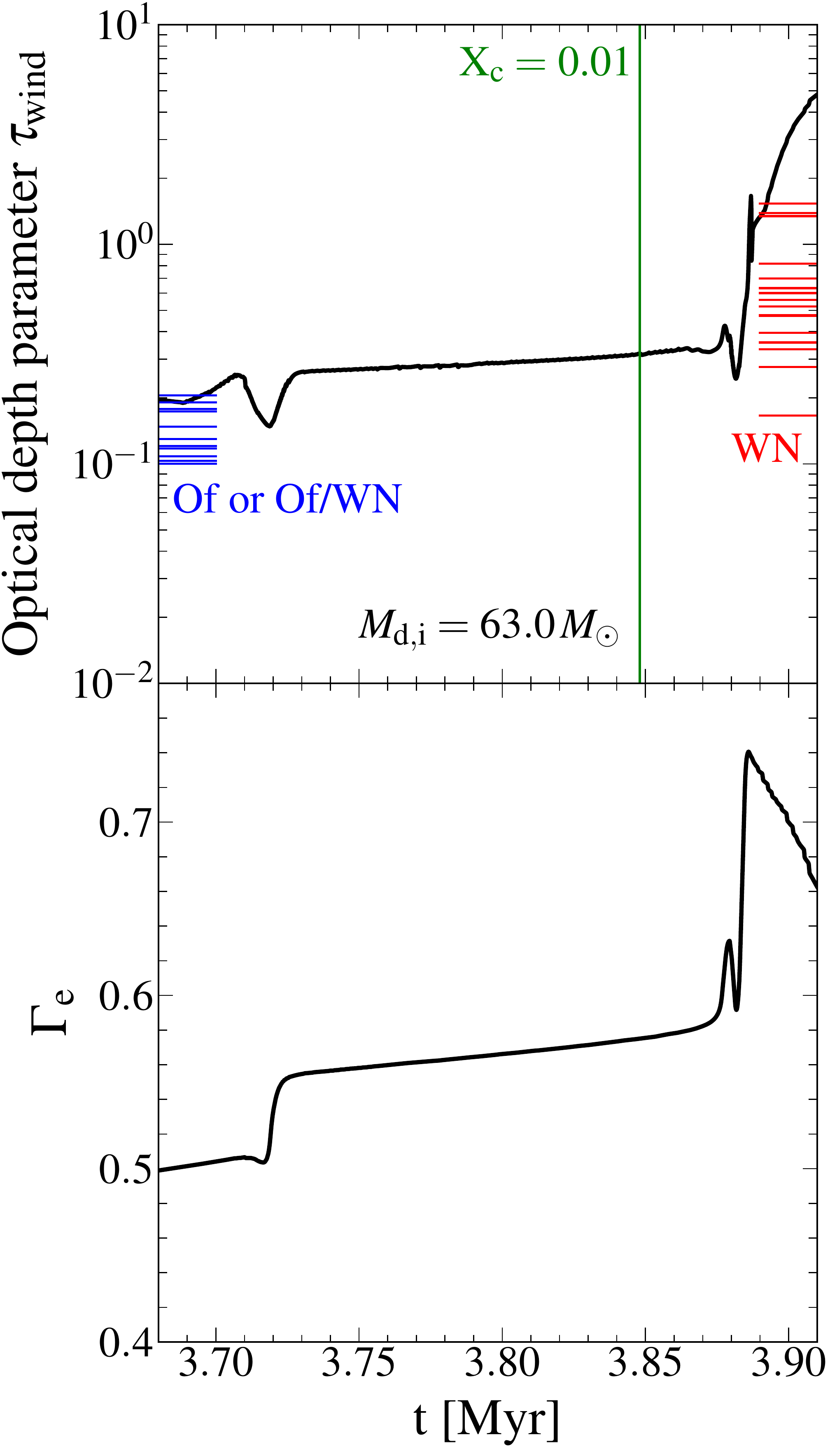}
    \includegraphics[width=0.48\hsize]{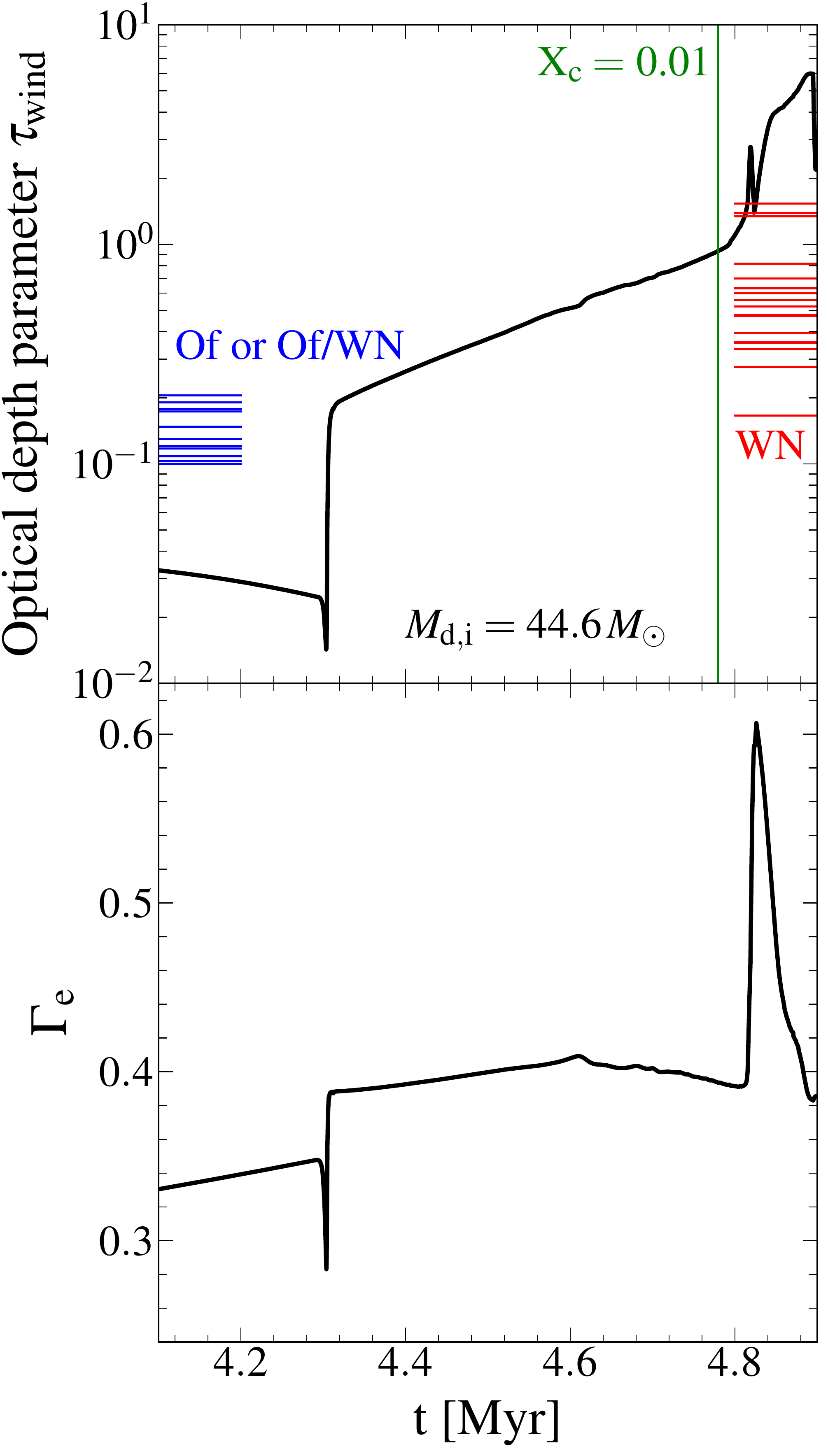}
    \caption{Time evolution of the optical depth parameter $\tau_{\rm wind}$ (\textit{top panels}) and the electron scattering Eddington factor $\Gamma_{\rm e}$ (\textit{bottom panels}, Eq.\,\ref{eq:GAMMA_E}) of the donor (black lines) for our two example models (Appendix.\,\ref{section_examples}). The X-axis limits are the same as in Fig.\,\ref{fig:de_reverse}. The green vertical lines indicate the age at which the central hydrogen mass fraction of the donor is equal to 0.01, that is, the donor completes core hydrogen burning. The red and blue horizontal lines denote the calculated optical depth parameter of the observed LMC WN stars and Of or Of/WN stars, respectively, which have a surface hydrogen mass fraction $X_{\rm H} \geq 0.4$ (Table\,\ref{table:tau_observations}). }
    \label{fig:tau}
\end{figure*}

\section{Comparison with observations}
\label{section_observations}

Here, we discuss several observed binary systems in which the 
reverse Algol channel has likely occurred. We compare detailed 
binary evolution models to observed binaries and look for 
indications that may support the reverse Algol scenario as such. 

\subsection{Reverse Algols in the Tarantula Nebula}
\label{obs1}

The recent Tarantula Massive Binary Monitoring (TMBM) program 
\citep{almeida2017,mahy2019a,mahy2019b} investigated the radial 
velocity variable O\,stars which were recognized as such as part of 
the VLT Flames Tarantula Survey (VFTS, \citealp{Evans2011,sana2013}).
As the Tarantula region in the LMC has been found to be particularly 
rich in very massive stars \citep{schneider2018}, it appears well 
suited to search for reverse Algol systems. In this sample, 
\cite{mahy2019a} identified five semi-detached binaries, roughly 
as expected from recent binary evolution models \citep{Sen2021b}.

From these five systems, \cite{mahy2019a} found two, VFTS\,094 and 
VFTS\,176, in which the Roche lobe filling component is the more 
massive star in the binary (see Sect.\,\ref{section_r} and panel `a' 
of Fig.\,\ref{fig:de_reverse}). While in principle, those could be 
systems caught during the fast Case\,A mass transfer and as such 
be progenitors of ordinary Algol systems, this is only expected 
for about 1 in one hundred semi-detached binaries. Finding two 
out of five semi-detached binaries with more massive donors argues 
strongly for all five systems evolving on the nuclear timescale, 
and thus for VFTS\,094 and VFTS\,176 being two genuine reverse 
Algol binaries.

The two mass donors in VFTS\,094 and VFTS\,176 have dynamical masses 
of 30.5\,$\mso$ and 28.3\,$\mso$, respectively. When we assume that 
these masses correspond to their initial convective core masses, their 
initial stellar masses are expected to be about $46\mso$ and $43\mso$, 
respectively (Fig.\,\ref{fig:inversion_pspace}). For 
non-conservative mass transfer, the initial masses 
of the accretors would be similar to their current masses, i.e., about 
$28\,\mso$ and $17\,\mso$, leading to initial mass ratios of $\sim0.6$ 
and $\sim0.4$, respectively. Figure\,\ref{fig:inversion_pspace} 
shows that under these assumptions, both systems are located inside 
the blue shaded area (large symbols), and their status as reverse Algol
is therefore consistent with our simple estimate. 

Assuming that conservative mass transfer leads to initial masses for the 
accretors of 13.0\,$\mso$ and 2.0\,$\mso$, and thus to initial mass 
ratios of 0.28 and 0.05, for VFTS\,094 and VFTS\,176, respectively. 
Conservative binary models in this parameter range are not available, 
but are unlikely to avoid a merging of the two stars during fast 
Case\,A mass transfer, since the thermal timescale ratio in massive 
main sequence binaries is roughly equal to $1/q^2$, 
that is, of order 10 for VFTS\,094, and about 160 for VFTS\,176. The 
mass gainers might therefore quickly swell up and lead to a contact 
configuration \citep{wellstein2001}. We conclude that certainly VFTS\,176, 
and likely also VFTS\,094, have evolved through a non-conservative 
fast Case\,A mass transfer phase. 

This expectation is confirmed when we compare with corresponding 
detailed binary evolution models of \citet{Pauli2022}, as we can 
see from Fig.\,\ref{fig:inversion_models}. For VFTS\,094, 
binary models with an initial mass ratio of 0.6 can undergo an 
ordinary or reverse Algol phase depending on the orbital period.
At the shortest orbital periods, tides prevent the spin-up of the accretor 
that leads to a mass transfer efficiency of up to 70\%. Therefore, 
the corresponding models in Fig.\,\ref{fig:inversion_models} develop 
only a short or no reverse Algol phase. Consequently, the observed 
masses of VFTS\,094 imply that the mass transfer efficiency in these 
models is overestimated. We conclude that while the models of 
\citet{Pauli2022} do not fit the main parameters of VFTS\,094, similar 
models with a reduced mass transfer efficiency have a high chance to 
reproduce this reverse Algol binary. 

For VFTS\,176, the initial mass ratio is more extreme.
At the shortest orbital periods \citep[see Fig.\,F2 of][]{Sen2021b}, 
the models shown in Fig.\,\ref{fig:inversion_models} undergo near-conservative 
mass transfer and do not show an extended reverse Algol phase. As for VFTS\,094, 
the implication is that the mass transfer efficiency in the corresponding 
models of \citet{Pauli2022} is significantly too high. A mass transfer 
efficiency of 70\% implies an initial mass ratio of $q_{\rm i}=0.17$ 
for VFTS\,176, while the short-period models in Fig.\,\ref{fig:inversion_models} 
avoid merging only for $q_{\rm i} \simgr 0.3$. Therefore models 
that might reproduce VFTS\,176 require a mass transfer efficiency 
which is much lower than 70\%. Again, such models have a high chance 
to reproduce this reverse Algol system, because for the lowest mass 
transfer efficiency, merging is always avoided.

The discussion above shows that VFTS\,094 and VFTS\,176 may be 
understood as reverse Algol systems. While current detailed models do 
not reproduce these two binaries, models which adopt less efficient 
mass transfer likely can. This raises the question of how Nature might 
achieve this. For the short orbital periods found in these two binaries, 
tides are expected to be efficient and to avoid a spin-up of the accretor 
to critical rotation, which is confirmed by their observed rotation 
rates \citep{mahy2019b}. Stellar rotation may therefore be of limited 
help for driving the excess mass out of the system during fast mass 
transfer, and the mass transfer scheme adopted in \citet{Pauli2022} 
needs to be revised for very massive short period binaries. Hence, 
these reverse Algol binaries provide a new pathway to constrain input 
physics assumptions in massive binary evolution. 

\subsection{Wolf-Rayet stars from the reverse Algol channel}
\label{obs2}

\begin{table*}
\centering
\caption{Stellar and optical depth parameter $\tau_{\rm wind}$ of binary and single WN stars in the LMC with a surface hydrogen mass fraction $X_{\rm H}$ $\geq$ 0.40.
}
\label{table:tau_observations}

\begin{tabular}{l c r r r r r r r r r r}
\hline
\hline \rule{0pt}{3ex}
BAT99 & Spectral type & log\,$L$      & $T_{*}$ & $R_{*}$           & $\upsilon_{\infty}$ & log\,$\dot{M}$   & $X_{\rm H}$ & $\tau_{\rm wind}$ & P$_{\rm orb}$ & $M_{\rm comp}/M_{\rm WR}$ \\ 
\#    &               & [$L_{\odot}$] & [kK]        & [$R_{\odot}$] & [km\,s$^{-1}$]      & [$M_{\odot}$ yr$^{-1}$] &             &        & [d]        &     \\
\hline
{\bf Binary} &        &               &             &               &                     &                  &             &        &               &     \\
006	  & O3 If*/WN7    & 5.90          & 45          & 15.0          & 1800                & $-$5.60            & 0.70        & 0.10   &          2.0  &     \\ 
072	  & WN4           & 5.05          & 70          &  2.3          & 1800                & $-$5.50            & 0.40        & 0.70   &               &     \\ %
077	  & WN7           & 5.79          & 45          & 13.0          & 1000                & $-$5.20	         & 0.70        & 0.47   &          3.0  & 1.66 $\pm$ 0.20    \\ 
079	  & WN7           & 5.90	      & 45	        & 15.0	        & 1000	              & $-$4.60	         & 0.40        & 1.34   &               &     \\ 
107	  & O6.5 Iafc     & 6.09	      & 33	        & 34.0          & 1300                & $-$5.20            & 0.70        & 0.15   &        153.9  & 0.81 $\pm$ 0.02    \\ 
113	  & O2 If*/WN5    & 6.14	      & 47	        & 18.0	        & 1800	              & $-$5.50            & 0.70        & 0.11   &          4.7  & 0.32 $\pm$ 0.04    \\ 
116$^{\rm a}$	  & WN5h          & 6.43	      & 53	        & 19.0	        & 2500	              & $-$4.20            & 0.65        & 1.53   &        154.5  & 0.92 $\pm$ 0.07    \\ 
119	  & WN6h          & 6.35	      & 50	        & 20.0	        & 1200	              & $-$4.40            & 0.40        & 1.38   &        158.7  & 1.01 $\pm$ 0.05    \\ 
\hline
{\bf Single} &        &               &             &               &                     &                  &             &        &               &     \\
012	  & O2 If*/WN5    & 5.80	      & 50	        & 10.6	        & 2400	              & $-$5.53            & 0.50        & 0.12   &               &     \\ 
068	  & O3.5 If*/WN7  & 6.00	      & 45	        & 16.7	        & 1000	              & $-$5.46	         & 0.60        & 0.19   &               &     \\ 
081	  & WN5h          & 5.48	      & 47	        &  8.2	        & 1000	              & $-$5.55	         & 0.40        & 0.27   &               &     \\ 
093	  & O3 If*        & 5.90          & 45          & 14.9	        & 1600	              & $-$5.63	         & 0.60        & 0.10   &               &     \\ 
097	  & O3.5 If*/WN7  & 6.30	      & 45	        & 23.7	        & 1600	              & $-$5.18	         & 0.60        & 0.18   &               &     \\ 
098	  & WN6           & 6.70	      & 45          & 37.5	        & 1600	              & $-$4.43	         & 0.60        & 0.63   &               &     \\ 
102	  & WN6           & 6.80	      & 45	        & 42.1	        & 1600	              & $-$4.21            & 0.40        & 0.82   &               &     \\ 
103	  & WN5(h)        & 6.25	      & 47	        & 19.9	        & 1600	              & $-$4.70	         & 0.40        & 0.56   &               &     \\ 
104	  & O2 If*/WN5    & 6.06	      & 63          &  9.0	        & 2400	              & $-$5.34	         & 0.40        & 0.20   &               &     \\ 
105	  & O2 If*        & 6.40	      & 50	        & 21.1	        & 1600	              & $-$5.41	         & 0.60        & 0.18   &               &     \\ 
106	  & WN5h          & 6.51	      & 56	        & 19.0         	& 2400	              & $-$4.55	         & 0.40        & 0.60   &               &     \\ 
108	  & WN5h          & 6.87	      & 56	        & 28.8	        & 2400	              & $-$4.43	         & 0.40        & 0.52   &               &     \\ 
109	  & WN5h          & 6.69	      & 56	        & 23.4	        & 2400	              & $-$4.56	         & 0.40        & 0.47   &               &     \\ 
110	  & O2 If*        & 6.22	      & 50	        & 17.1	        & 2400	              & $-$5.22	         & 0.70        & 0.17   &               &     \\ 
111	  & WN9ha         & 6.25	      & 45	        & 22.3	        & 1000	              & $-$5.42	         & 0.70        & 0.17   &               &     \\ 
114	  & O2 If*/WN5    & 6.44	      & 63          & 13.9	        & 2400	              & $-$5.35	         & 0.40        & 0.13   &               &     \\ 
117	  & WN5ha         & 6.40	      & 63          & 13.3	        & 2400	              & $-$4.93	         & 0.40        & 0.36   &               &     \\ 
130	  & WN11h         & 5.68	      & 28	        & 29.1	        & 200                 & $-$5.35	         & 0.40        & 0.40   &               &     \\ 
133	  & WN11h         & 5.69	      & 28	        & 29.4	        & 200                 & $-$5.42	         & 0.40        & 0.33   &               &     \\ 
\hline
\hline
\end{tabular}
\newline
\newline
\tablefoot{The stellar parameters of the binary and single WN (and O) stars are taken from \citet{Shenar2019,Shenar2020corr} and \citet{Hainich2014} respectively. The stellar bolometric luminosity, temperature and radius are given by $L$, $T_{*}$ and $R_{*}$. The terminal wind-speed and surface hydrogen mass fraction is given by $\upsilon_{\infty}$ and $X_{\rm H}$. The wind mass-loss rate of the WN star is given by $\dot{M}$. The orbital period and mass ratio (mass of companion divided by mass of WN star) of the binary systems are given by $P_{\rm orb}$ and $q$ respectively, whenever available \citep[][]{Shenar2019}. We quote the mass ratios for the WR binaries for which \citet{Shenar2019} found SB2 solutions, but not from the spectroscopic masses reported in Table\,2 of \citet{Shenar2019}. (a) Parameters adopted from \citet{Tehrani2019}. 
}
\end{table*}

As detailed in Sect.\,\ref{section_wr}, we expect the donor stars of Case\,A
binaries after the fast Case\,A mass transfer to be overluminous.
As such, their stellar wind mass loss rate will be elevated compared to
single stars of the same luminosity. Since the donor's surface may still be un-enriched in helium at this stage, we expect O stars (evolving as single stars)
and WR stars (Algol donors), both hydrogen-rich, at the same position
in the HR diagram (Fig.\,\ref{fig:hrds}). 

The stripping of the donor star due to Roche-lobe Overflow will occur 
in essentially the same way in ordinary and reverse Algol binaries. In 
reverse Algols, however, the donor, or the WR component, may outshine 
the companion more easily, because it is more massive and has the higher 
average mean molecular weight, which both contribute to make it the more 
luminous of the two stars. When the mass ratio is more extreme, the 
companion star may not be detected at all, and the WR component could 
be misinterpreted as a hydrogen-rich single WR star. 

The majority of the observed WR stars are expected to undergo core helium 
burning \citep{Pauli2022}. However, the analysis of luminous hydrogen-rich 
WN stars in the LMC and the Galaxy showed some of them to likely be core 
hydrogen burning stars \citep{deKoter1997,Martins2013}. 
To compare our donor models with observed WR stars, one needs a 
quantitative criterion allowing us to assess whether a given model 
corresponds to a WR star or not. For this purpose, we consider 
the optical depth parameter introduced in
Sect.\,\ref{subsection_optical_depth_method}. 

Figure\,\ref{fig:tau} (top panels) shows the evolution of the optical 
depth parameter $\tau_{\rm wind}$ (Eq.\,\ref{eqn:tau}) for the donor 
stars of our two example binary models (Appendix\,\ref{section_examples}). 
For the more massive model (left 
panel), we find $\tau_{\rm wind} \simeq 0.2$ before any mass is transferred. 
This corresponds to a high wind mass-loss rate of more than $\sim$10$^{-5}$\,$M_{\odot}$\,yr$^{-1}$ 
shortly before the onset of mass transfer (Fig.\,\ref{fig:sd_reverse}). 
After the fast Case\,A mass transfer phase, the wind mass-loss rate 
increases slightly, because about 10\,$M_{\odot}$ are removed 
from the donor. This in turn only leads to an increase of the donor's 
optical depth parameter to $\sim$0.3. Since the radius of the 
donor star is slowly increasing (Fig\,\ref{fig:sd_reverse}),
this binary model remains in the semi-detached configuration while the 
orbital period is also slowly increasing.
Likewise, the wind mass-loss rate 
is also increasing during this slow Case\,A mass transfer phase. 
These two effects compensate each other such that the optical depth parameter 
of the donor star remains nearly constant (see Eq.\,\ref{eqn:tau}) 
until the end of core hydrogen burning. 
After core hydrogen depletion and the thermal timescale Case\,AB mass transfer, the wind mass-loss rate 
increases again and the donor radius shrinks drastically, leading to a 
sharp rise of the optical depth parameter. 

For the less massive model (right panel), the optical depth parameter 
of the donor is very low, $\sim$0.03, before the onset of mass transfer 
as its wind mass-loss rate is more than on order of magnitude smaller
than that of the model discussed above. 
During fast Case\,A mass transfer, both the wind mass-loss rate 
and the optical depth parameter of the donor 
rise by almost one order of magnitude (Fig.\,\ref{fig:de_reverse}).
The optical depth parameter grows to a value near unity 
during the remainder of the core hydrogen burning evolution
of the donor, and to even higher values afterwards. 

In order to deduce from the optical depth parameters of the models
whether they may correspond to WR stars,
we calculate the optical depth parameters of observed  
hydrogen-rich WN and Of stars in the LMC. 
For this we use the observed stellar radius, terminal wind velocity, and wind 
mass-loss rate in Eq.\,\ref{eqn:tau}.  Table\,\ref{table:tau_observations} shows 
that the optical depth parameters of the Of and 
Of/WN stars fall in the range $\tau_{\rm wind}= 0.10 \dots 0.20$,
while those of the remaining WN stars cover $\tau_{\rm wind}= 0.17 \dots 1.53$. 
Therefore it seems reasonable to adopt a threshold of $\tau_{\rm wind} \simgr 0.2$ 
for assuming that a given donor model represents a 
hydrogen-rich Wolf-Rayet star.

In doing so, we see from the more massive binary model in
Fig.\,\ref{fig:tau} that the donor may already develop WR characteristics
before mass transfer starts. Therefore, this would equally apply to
a single star with the same initial mass. This consideration shows that
above a certain initial mass limit, single stars may become
WR stars during core hydrogen burning, as has anticipated before 
in the literature \citep[e.g.][]{shenar2020}. According to our example,
this initial mass limit could be just above $60\mso$ in the LMC.
However, for smaller mass loss rates \citep{Bjorklund2021,
Hawcroft2021,Brands2022}, the limiting mass may be significantly larger.

Our second example binary shows that, for an initial mass of 
$\sim 45\mso$, the donor star will not show WR characteristics 
before the mass transfer. However, Fig.\,\ref{fig:tau} suggests that, 
in this case, the elevated L/M-ratio after fast Case\,A mass transfer 
pushes the donor into the Wolf-Rayet regime, and that during its 
reverse Algol-like phase the WR characteristics becomes stronger with time. 
Here, the donor has a mass which would clearly not lead it to become 
a hydrogen burning WR star if it were a single star, but in a binary system, 
it evolves into a $\sim 28\mso$ hydrogen rich, core hydrogen burning WN star. 


In previous works the occurrence of WR-type winds has been
attributed to the proximity of the WR stars to their Eddington limit
\citep{grafener2008,grafener2011,sander2020}. For very massive
main-sequence stars in the LMC, \citet{bestenlehner2014} identified
a transition between optically-thin O\,star winds and enhanced WR-type
winds at electron scattering Eddington factors of $\Gamma_{\rm e} \approx 0.4$. 
The post-case\,A donors in our example models show similarly high values of 
$\Gamma_{\rm e}$ of about 0.4 and 0.5 (bottom panels of Fig.\,\ref{fig:tau}), 
supporting the interpretation that they represent WR stars.

After this phase, for a short period after the thermal timescale Case\,AB mass transfer (see 
Panel\,b of Fig.\,\ref{fig:de_reverse}), The Eddington factor increases
even further, to values of 0.6 and 0.7 respectively. Finally, when the core 
He-burning WR phase is reached, it settles back down to values in the range 
of 0.4 to 0.5. Again, that the Eddington factors of the models during the 
reverse Algol phase are similar to those of the same models during core helium 
burning --- where they clearly correspond to hydrogen-poor WR stars \citep{Pauli2022} 
--- supports the WR interpretation during their reverse Algol phase. 

\subsection{Observed counterpart Wolf-Rayet stars}

With their moderate hydrogen deficiencies and low effective temperatures, 
the models described above are reminiscent of H-rich WNL stars as 
those observed in the LMC sample of \citet{Hainich2014}, at luminosities 
of $\log(L/L_\odot)\approx 5.5\dots5.9$ (see their Fig.\,8). These objects 
have luminosities comparable to classical WR stars, but show much lower 
mass-loss rates (Fig.\,6 in \citealp{Hainich2014}), indicating a distinct 
physical nature from classical WR stars.


Table\,\ref{table:tau_observations} lists all hydrogen-rich WR star 
in the LMC ($X_{\rm H} \geq 0.4$). The orbital periods of the 
detected binaries are either smaller than 5\,d or larger than 150\,d. 
Since it appears unlikely that no WN binary with an orbital period 
in the range 5$\dots$150\,d exists in the LMC (see, for e.g., 
\citealp{langer2020}), some of the apparently single WN stars from 
this list may also have main sequence companions. That they are not 
detected implies that they are significantly less luminous than the 
WR stars, which means they could be counterparts of our post-mass 
transfer Case A models in which the primary star remains more luminous 
or even more massive (Sect.\,\ref{section_r}). 

The three very luminous WN stars in long-period binaries listed in
Table\,\ref{table:tau_observations} could just be wind stripped, i.e., 
their initial mass might be larger than $M_{\rm WNH}$ defined in 
Sect.\,\ref{section_wr}. This is particularly true for BAT99\,\#116 
(see also, \citealp{Tehrani2019}) and BAT99\,\#119 (see also, 
\citealp{Shenar2017}), which contain very luminous WR stars. However, 
due to the effect of envelope inflation binaries with initial orbital 
periods of up to $\sim$2000\,d can undergo Case\,A mass transfer 
(Sect.\,\ref{section_A}), such that an Algol evolution can not be 
excluded for BAT99\#107. 

Of the short-period binaries listed in Table\,\ref{table:tau_observations},
BAT99\#113 is the only one in which the WR component is known to be 
significantly more massive than the companion. The WR component appears 
to be close to Roche lobe filling, and its high hydrogen abundance 
excludes a previous Case\,B mass transfer. Using the luminosity and
dynamical mass estimate for the WR component of \citet{Shenar2019} 
yields a luminosity-to-mass ratio of $\log{\mathscr L}/ {\mathscr L}_{\odot} = 4.41$ 
with ${\mathscr L} = (4\pi\sigma G)^{-1} {L / M}$. Notably, this value 
is only about 0.2\,dex from the electron-scattering Eddington limit 
($\log{\mathscr L}/ {\mathscr L}_{\odot} = 4.6$, \citealp{Langer2014}), 
whereas corresponding single star models as hot as the WR star in 
BAT99\#113 ($T_{\rm eff}\simeq 45\,$kK) remain below $\log{\mathscr L}/ {\mathscr L}_{\odot} = 4.2$ 
(see Fig.\,18 of \citealp{Kohler2015}). 

In the detailed binary evolution grid, the model with the initial 
parameters $(M_{1, \rm i},M_{2, \rm i},P_{\rm i})=(79.4\mso,19.8\mso,15.8\,{\rm d})$
obtains, after fast Case\,A mass transfer, $(M_{\rm 1},M_{\rm 2},P)=53\mso,20\mso,4.5\,{\rm d})$,
which fits well to the values derived by \citet{Shenar2019} of
$(M_{\rm WR},M_{\rm OB},P)=52^{+20}_{-15}\mso,17^{+2}_{-2}\mso,4.7\,{\rm d})$.
At an effective temperature of $43\,$kK, the donor's luminosity-to-mass ratio is
$\log{\mathscr L}/ {\mathscr L}_{\odot} = 4.37$. A merging of both 
stars is avoided. We conclude that BAT99\#113 constitutes a strong 
case for reverse Algol evolution. This argues for stable mass transfer 
in very massive binaries even for rather extreme initial mass ratios 
(here $q_{\rm i}=0.25$) due to highly non-conservative mass transfer. 



\begin{figure}
    \centering
    \includegraphics[width=\hsize]{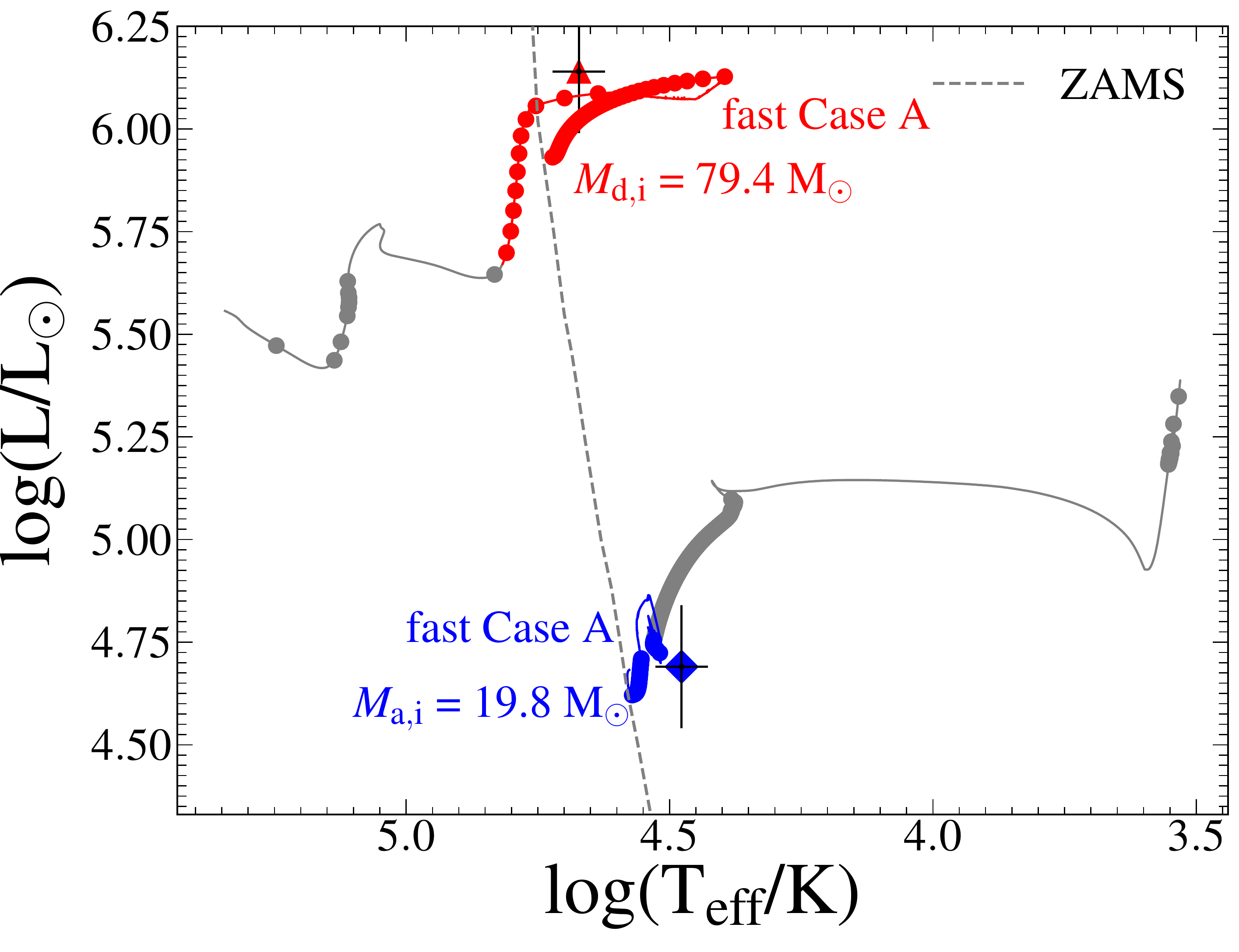}
    \caption{HR diagram showing the evolution of luminosity and surface temperature of donor and accretor star from a binary evolution model with initial donor ($M_{\rm d,i}$) and accretor masses ($M_{\rm a,i}$) of 79.4$\mso$ and 19.8$\mso$, respectively, and an initial orbital period of 15.8\,d. Dots are placed along the tracks every 50\,000 years. Colored lines and symbols (red for the donor and blue for the accretor) are used as long as the donor undergoes core hydrogen burning. The remaining evolution of both stars is shown in gray. The zero age main sequence is shown as dashed gray line. The red triangle and blue diamond indicate the position of the WR star and its companion in BAT99\#113 respectively, with their error bars. }
    \label{fig:hrd_BAT99_113}
\end{figure}

Figure\,\ref{fig:hrd_BAT99_113} shows the evolution of 
luminosity and effective temperature of both components 
of the binary model mentioned above. 
When the more luminous donor fills its Roche volume 
at the coolest point of the track, fast Case\,A mass transfer is initiated, and the reverse Algol phase stars immediately 
thereafter.
The donor remains more massive and more luminous than the accretor, in contrast to the ordinary Algol evolution in less massive binaries \citep{pols1994,nelson2001,Sen2021b}. 
Since the mass transfer efficiency in this model is low, the 
accretor evolves similar to a single star of the corresponding 
mass. The HRD positions of both model components
after the fast Case\,A mass transfer match the observed 
positions of the WR star and its companion in BAT99\#113. 

We note that \citet{Shenar2019} attempted to derive the initial 
parameters and evolutionary state of the WR binaries listed in 
Table\,\ref{table:tau_observations} by comparing with the binary 
evolution grid of \citet{Eldridge2008,Eldridge2016}. 
However, for BAT99\#113, no satisfactory fit could be obtained. 

\section{Discussion and uncertainties}
\label{section_discussion}

\subsection{Envelope inflation}
\label{section_inf}

\cite{sanyal2015,sanyal2017} studied the detailed single star models 
of \citet{brott2011} and \citet{Kohler2015} and found that models 
exceeding $\sim$40 $M_{\odot}$ at LMC metallicity reach their Eddington 
limit inside the stellar envelope towards the end of core hydrogen 
burning. The stellar envelope expands and the donors grow to red 
supergiant proportions towards core hydrogen exhaustion. Due to this 
inflation, \citet{Pauli2022} found that binary models with an initial 
donor mass of $\sim$50$\mso$ and $\sim$56.2$\mso$ can undergo Case\,A 
mass transfer for initial orbital periods up to 120\,d, and 2000\,d 
respectively. This leads to a very large parameter space for reverse 
Algol evolution in very massive binaries. 

However, the extent of the envelope inflation in 1D models depends on
the mixing length parameter \citep{sanyal2015}, or, more generally, on 
the employed model for convective energy transport. For more efficient 
convection, inflation is reduced, and so would be the orbital period 
range for Case\,A binaries above $\sim$40 $M_{\odot}$. In this case, 
the mass range for reverse Algol evolution would remain the same, but 
the number of systems experiencing it would be reduced. Notably, in the 
Case\,B binaries which would instead be produced, the donor star would 
be significantly less massive and more helium-rich than a comparable 
Case\,A binary. 


\subsection{Mass transfer efficiency and stability of mass transfer}
\label{sec:stability}
Recently, \citet{Sen2021b} showed that an accretor 
spin dependent mass transfer efficiency can explain many 
observed massive Algol binaries reasonably well. Observations 
of individual binary systems also indicate that the mass transfer 
efficiency can vary from binary to binary. While some binaries favour 
low mass transfer efficiency \citep{langer2003}, others indicate a 
need for higher mass transfer efficiency \citep{wellstein1999}. 
\citet{petrovic2005} and \citet{selma2007} also found evidence for 
a mass ratio and orbital period-dependent mass transfer efficiency, 
respectively. 

The accretor spin dependent mass transfer efficiency 
results in an orbital period and mass ratio 
dependence of the mass transfer efficiency \citep{langer2020,Sen2021b}. 
As already discussed, inefficient mass transfer 
leads to more massive binaries being able to 
undergo the reverse Algol evolution. When the mass transfer 
efficiency decreases with increasing orbital period, we found 
that the reverse Algol evolution also shows an orbital 
period dependence (see Fig.\,\ref{fig:inversion_models}). For a fixed 
initial mass ratio, lower orbital period models are more 
likely to invert their mass ratio during fast Case\,A mass 
transfer and vice versa. 

\citet{Pauli2022} determine the stability of their binary models 
during inefficient mass transfer by assuming that the combined 
photon energy from both the binary components is larger 
than the gravitational energy needed to remove the excess 
mass transferred when the accretor reaches critical rotation.  
As discussed in Sect\,\ref{obs1}, the two reverse Algol binaries 
in the Tarantula region require less efficient mass transfer 
than provided by this mechanism. 

The mass transfer efficiency is also affecting the
state of double black hole systems. If the accretion efficiency is small,
then the initially more massive star is expected to produce the more massive 
black hole (e.g., Fig.\,7 of \citealp{langer2020}). For conservative evolution, 
however, this is so only in the binaries with the lowest initial mass ratios,
while in many systems the initially more massive star becomes the less massive one
after mass transfer (e.g., in the classical Algol evolution), and may form
the lower mass black hole. As the spin parameters of the first and second formed
black holes are potentially different, the measured spins in black hole mergers may 
have an inference on the mass transfer efficiency in massive binaries \citep{Mould2022}, 
and thus on the fraction of very massive binaries undergoing the reverse Algol evolution.

\subsection{Wind mass-loss rates}

It has been shown that the WR 
mass-loss rates given by \citet{Nugis2000} are not adequate to 
explain the luminosity distribution of WC- and WO-type stars and 
the observed properties of SN Ic progenitors \citep{yoon2017b}. 
It was also recommended to use a clumping factor of $D = 4$ instead 
of the usual $D = 10$ that is compatible with 
the \citet{Nugis2000} prescription, to reproduce the 
distribution of galactic WN and WC stars \citep[see also][]{Pauli2022}. 
Quantitatively, using a lower clumping factor in the 
\citet{Nugis2000} prescription increases the wind mass-loss rates 
of mildly helium-enriched stars. 

The models used here include this updated wind-mass loss prescription of 
\citet{yoon2017b} with a clumping factor of D = 3 \citep[see][]{Pauli2022}. 
Hence, the wind mass-loss rates of our stripped donors on the 
main sequence have a greater wind mass-loss rate than previous 
models in the literature. This makes our donors
more likely to develop an optically thick wind after fast Case A
mass transfer (see Eq.\,\ref{eqn:tau}). 

Recent theoretical and observational studies indicate 
that the work of \citet{Vink2000} may overestimate the wind mass-loss 
rate of O type stars by a factor of 2-3 \citep{Bjorklund2021,
Hawcroft2021,Brands2022}. A lower mass-loss rate for O stars 
mean that it will be harder for single stars to get wind-stripped 
and show a WR-like spectrum. A lower wind mass-loss rate would 
also reduce the optical depth parameter just after the end of 
the fast Case\,A mass transfer phase. The jump in the wind 
mass-loss rate of the donor stars after fast Case\,A mass 
transfer due to an increased Eddington factor will still 
remain. Moreover, since the surface hydrogen mass fraction decreases 
below 0.7 shortly after the fast Case\,A phase, when the mass-loss 
rate is calculated by interpolating between the \citet{Vink2000} 
and \citet{Nugis2000} rates, we expect that the effect of a lower 
wind mass-loss rate of O stars will not affect our results 
significantly during the majority of the reverse Algol phase.

\subsection{Mass of the convective core}

The overshooting parameter in the models was calibrated to 
$\alpha_{\rm ov} = 0.335$ \citep{brott2011} against observations 
of the rotational velocities of massive stars in the FLAMES Survey 
of Massive Stars \citep{Hunter2008}, at masses around 16$\mso$. 
This value was confirmed by \citet{Castro2014}, who compared their 
sample of Galactic massive stars to single star models. However, 
they found that while this value of the overshooting parameter works 
for stars of mass $\sim$16$\mso$, a smaller or larger value is 
preferred for stars that are less or more massive than $\sim$16$\mso$, 
respectively. A more recent study by \citet{Castro2018} on SMC OB 
field stars \citep{Lamb2016} found that an overshooting parameter 
of 0.335 was able to reproduce the derived tentative TAMS line of 
these massive stars \citep[see also][]{Gilkis2021}. 

Recent observations of massive stars, via gravity mode asteroseismology 
of B stars \citep{Pedersen2021} and eclipsing binary systems \citep{Claret2019}, 
have indicated that their convective cores are larger 
than what is adopted in many stellar evolution models \citep{Maeder1988,Alongi1993}. 
Following this empirical evidence \citep[see also][]{Tkachenko2020}, 
theoretical studies have shown that convective 
penetration in early-type stars can increase the size of the convective 
core by $\sim$10\%-30\% of the pressure scale height 
at the core boundary \citep{Anders2022,Jermyn2022}.  A larger increase 
of the convective core size will lead to smaller envelope masses, 
therefore increasing the likelihood of reverse Algol evolution in 
even lower mass binaries. A quantification of this effect is 
however outside the scope of this work. 

\subsection{Metallicity}
\label{subsection_metallicity}

Wind mass-loss from hot massive stars \citep[for a review, see][]
{Puls2008} is known to depend on their metallicity \citep[for e.g., 
see][]{mokiem2007,Vink2000,Vink2021,Bjorklund2021}. In the LMC, 
the initial mass threshold for single stars to show WR-like winds 
is around $60\mso$. We showed that binary stripping can increase 
the luminosity-to-mass ratio of the donor such that stripped 
stars with masses above $40\mso$ can develop a WR-like wind spectrum. 
For lower (higher) metallicity environments, the wind mass-loss 
rates are lower (higher) as well, decreasing (increasing) the 
mass threshold for ordinary single stars to exhibit the WR phenomenon
\citep{shenar2020}. 

For binary-stripped donors, the wind mass-loss rate increase 
due to the enhanced luminosity-to-mass ratio occurs independent of the metallicity \citep{Vink2000}, with the mass-loss rate of a binary stripped donor being higher 
than that of a single star of the same mass. The mass threshold 
for binary stripped donors to show a WR-type spectrum also decreases 
(increases) for higher (lower) metallicity, with the 
absolute value at each metallicity being lower for the case of 
binary stripping than for the single star scenario. As also discussed 
in the literature \citep{shenar2020}, we expect a range of masses 
(and in turn luminosities) where only binary stripping can 
produce these hydrogen-rich WR stars, at each metallicity. 

\section{Conclusions}
\label{section_conclusion}

Semi-detached binary systems serve as an excellent test-bed for 
binary evolution and the involved physical processes \citep{selma2007,mennekens2017,Sen2021b}. 
The existence of a nuclear timescale mass transfer phase not only 
enables us to observe mass transferring binaries to which we can 
compare the models, but also gives us a window to understand 
the elusive but dominant thermal timescale mass transfer. 

In this work, we studied detailed binary evolution models with 
initial donor masses above $40\mso$. We focus on the configuration 
of semi-detached binaries in which the Roche-lobe filling donor star 
performs nuclear timescale mass transfer as the more 
massive component of the binary (Fig\,\ref{fig:sd_reverse}). 
This reverse Algol evolutionary channel in very 
massive binaries occurs due to the higher ratio of core to 
envelope mass in more massive stars. Provided the initial mass 
ratio of the binary is small enough (Fig.\,\ref{fig:inversion_pspace}), 
the envelope stripping of the donor does not remove enough mass 
to invert the mass ratio of the binary. We find that these 
stripped donors may remain hydrogen-rich and are highly overluminous 
for their mass. 

We identify two massive semi-detached binaries VFTS\,094 and 
VFTS\,176 \citep{mahy2019b,mahy2019a}, as likely observational 
counterparts of reverse Algol systems. However, their properties 
are shown to be incompatible with conservative fast Case\,A mass transfer,
and require an even lower mass transfer efficiency than what is provided
by the detailed binary evolution models of \citet{Pauli2022}.

In sufficiently massive binaries, the overluminosity of the donor 
may induce an elevated stellar wind mass-loss, leading to optical 
depth parameters comparable to those of the observed hydrogen-rich 
WN stars (Fig\,\ref{fig:tau}). We show that the WR binary BAT99\#113 
\citep{Shenar2019} is well explained by the reverse 
Algol scenario. We also identify several other LMC WR binaries as potential 
reverse Algol counterparts, and argue that some of the apparently 
single hydrogen-rich WR stars in the LMC (Table\,\ref{table:tau_observations}) 
might also currently be in this phase. 

Some very massive Case\,A binaries may be progenitors of high-mass 
black hole binaries and of double black hole systems. A comprehensive 
study of corresponding models grids is required to develop a deeper 
understanding of their evolution. Population synthesis based on rapid 
binary evolution models will have difficulties to achieve this, due to 
the complexity of Case\,A mass transfer. Besides more detailed evolutionary 
calculations for very massive binaries, refined models for fast mass 
transfer need to be developed \citep{Dessart2003,Lu2022} and tested 
against the observed Algol binary population. 


\begin{acknowledgements}
We thank the anonymous referee for helpful comments that improved the manuscript. This research was funded in part by the National Science Center (NCN), Poland under grant number OPUS 2021/41/B/ST9/00757. TS acknowledges support from the European Union's Horizon 2020 under the Marie Skłodowska-Curie grant agreement No 101024605. The research leading to these results has received funding from the European Research Council (ERC) under the European Union’s Horizon 2020 research and innovation programme (grant agreement number 772225: MULTIPLES).

\end{acknowledgements}

\bibliographystyle{aa}
\bibliography{caseA}

\begin{appendix}

\section{Examples of reverse Algol evolution models}
\label{section_examples}

In this section, we show two examples of Case\,A mass transfer for very 
massive binary models and discuss their differences compared to the typical 
nuclear timescale Case\,A mass transfer phase studied in the literature 
\citep{nelson2001,selma2007,Sen2021b}. 

\begin{figure*}
    \centering
    \includegraphics[width=0.48\hsize]{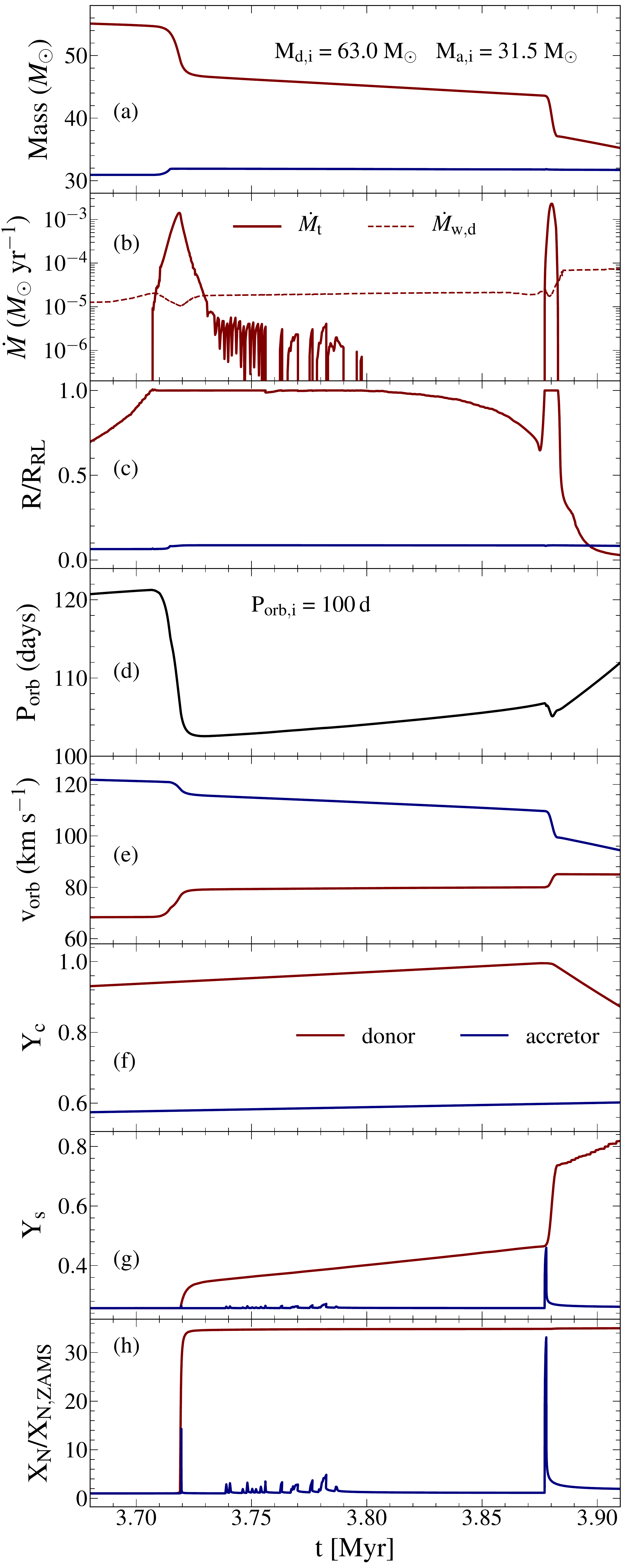}
    \includegraphics[width=0.48\hsize]{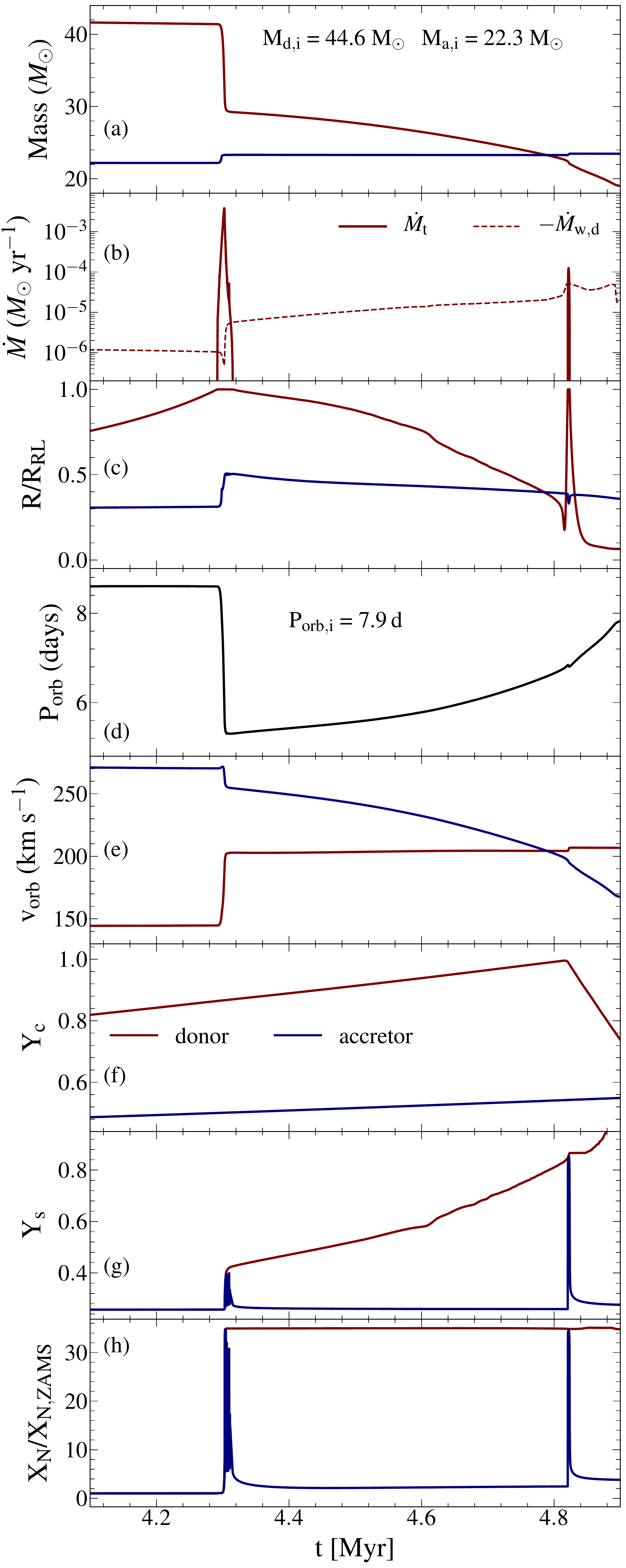}
    \caption{Examples of reverse Algol evolution. \textit{Left panel:} A model that undergoes nuclear timescale slow Case\,A mass transfer in the reverse Algol configuration. The initial donor mass, mass ratio and orbital period of the model are 63.0  $M_{\odot}$, 0.5 and 100\,d respectively. \textit{Right panel:} A model that detaches after fast Case\,A mass transfer. The initial donor mass, mass ratio and orbital period of the model are 44.6  $M_{\odot}$, 0.5 and 7.9\,d respectively. We show selected stellar parameters as function of time, where $t=0$ denotes the ZAMS of both stars. \textit{(a):} Donor (red) and accretor (blue) mass. \textit{(b):} Mass transfer rate $\dot{M}_{\rm t}$ (thick red line) and wind mass-loss rate of the donor ($-\dot{M}_{\rm w,d}$, red dotted line). \textit{(c):} Ratio of the donor and accretor radius to their respective Roche lobe radius. \textit{(d):} Orbital period. \textit{(e):} Orbital velocity of donor and accretor. \textit{(f):} Central helium mass fraction of donor and accretor. \textit{(g):} Surface helium mass fraction. \textit{(h):} Surface nitrogen mass fraction enhancement factor.  
    }
    \label{fig:sd_reverse}
    \label{fig:de_reverse}
\end{figure*}

\begin{figure*}
    \centering
    \includegraphics[width=0.48\hsize]{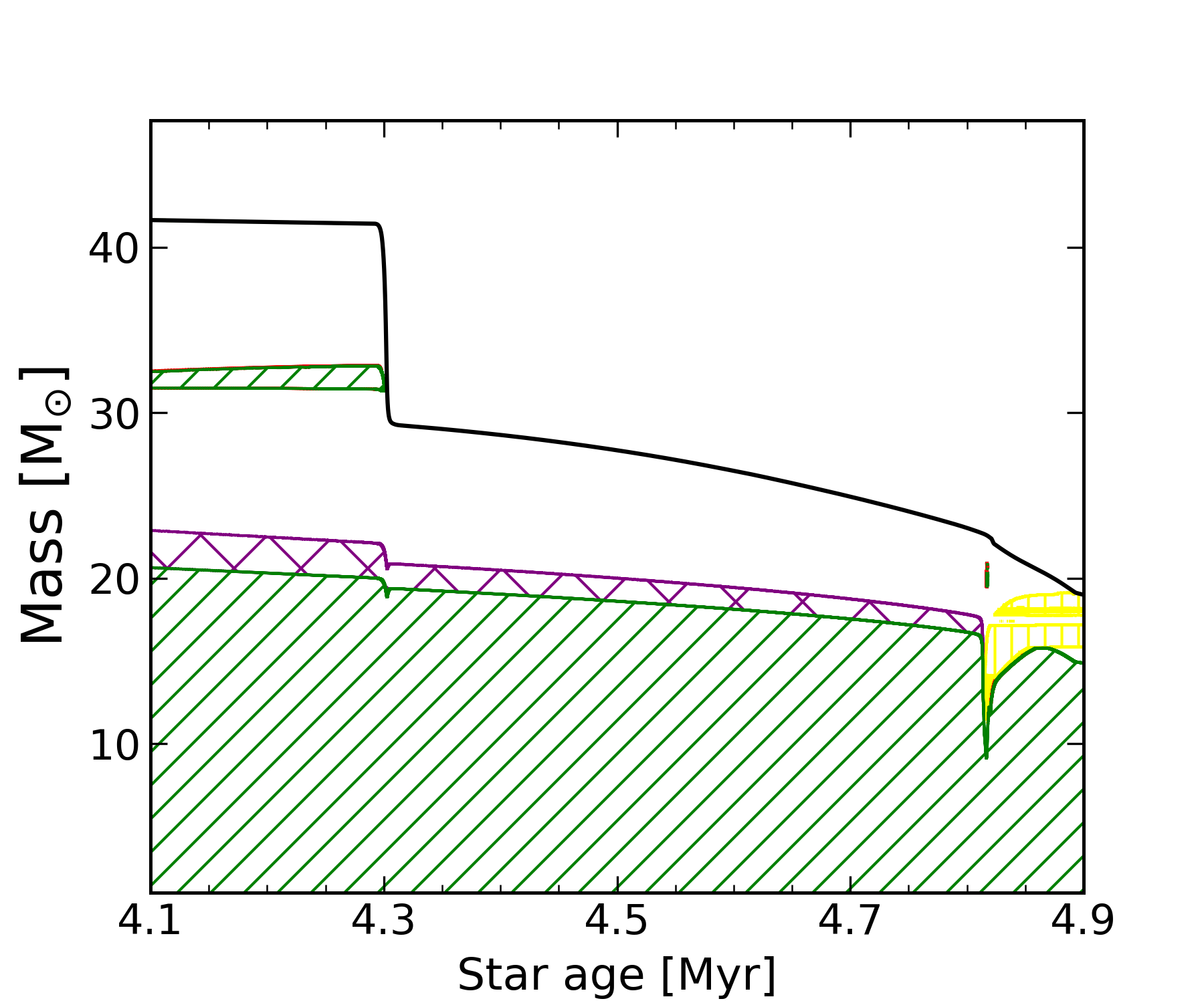}
    \includegraphics[width=0.48\hsize]{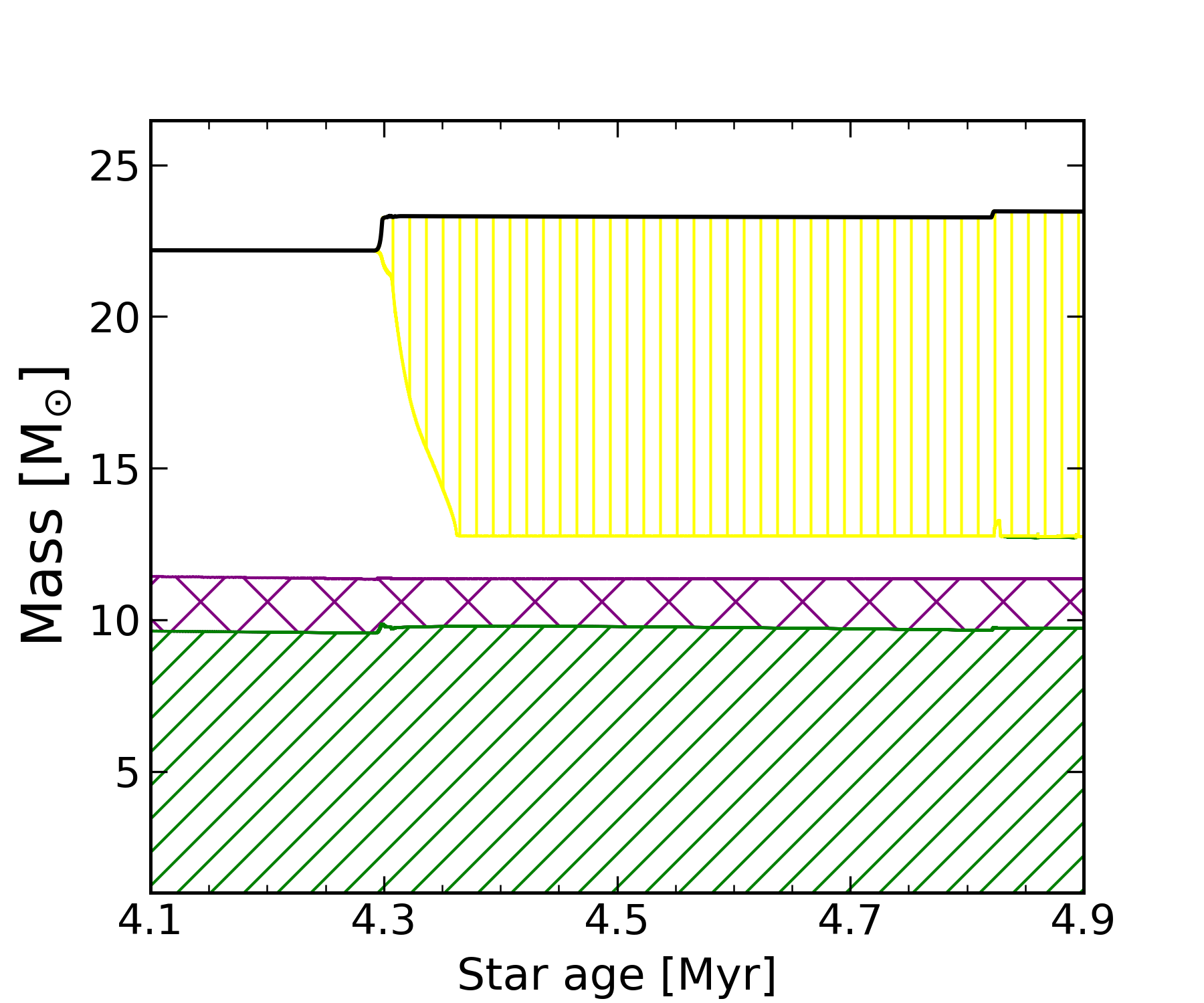}
    \caption{Kippenhahn diagram showing the internal structure of the donor and accretor of the model with initial donor mass, mass ratio and orbital period of 44.6  $M_{\odot}$, 0.5 and 7.9\,d respectively, as a function of the stellar age. The green hatched line shows the regions of the star that are unstable to convection. The purple cross hatched line shows the extent of convective overshooting above the convective core. The yellow hatching denotes the regions of thermohaline mixing. The red colour shows the regions of semi-convective mixing. The black line gives the total mass of the star as a function of stellar age. }
    \label{fig:kipp}
\end{figure*}

\subsection{Mass and mass transfer efficiency}
\label{subsection_sd}
\label{subsection_de}

Figure\,\ref{fig:sd_reverse} (left panel) shows the evolution of a massive 
binary model with a donor of initial mass of 63\,$\mso$ in a 100\,d orbit 
around an accretor of initial mass 31.5\,$\mso$. While both the binary 
components are burning hydrogen at their cores (Panel\,`f'), the donor star 
(red line) fills its Roche lobe (Panel\,`c') at around 3.71\,Myr, thereby 
commencing the Case\,A mass transfer phase. The mass transfer rate (thick 
red line) rises to about 10$^{-3}$ $M_{\odot}$\,yr$^{-1}$ (Panel\,`b') during 
the so-called fast Case\,A mass transfer phase, that is, of the order of 
$M_{\rm d}/\tau_{\rm KH}$, where $M_{\rm d}$ is the mass of the donor and 
$\tau_{\rm KH}$ is the thermal timescale of the donor at the onset of mass 
transfer. 

The fast Case\,A mass transfer continues at the thermal timescale because 
the mass transfer occurs while the binary is going towards a mass ratio of 
unity and is hence decreasing its orbital separation. When the envelope 
stripping reaches the mass coordinate of the initial convective core however, 
the mass-radius exponent in the hydrogen-helium gradient region is different 
such that the radius of the donor now only increases at the nuclear timescale. 
Then the system enters a nuclear timescale mass transfer phase at around 
3.73\,Myr. The donor loses most of its mass during the thermal timescale 
mass transfer phase (Panel\,`d'), while very little mass is lost via Roche-lobe 
overflow during the slow Case\,A mass transfer phase. 

Contrary to the usual Algol evolution \citep{Sen2021b}, we see that the 
mass ratio of the binary is not inverted during the entire Case\,A 
mass transfer phase (Panel\,`a'). In this configuration, the currently 
more massive donor is transferring mass to its less massive companion on 
the nuclear timescale. This constitutes the reverse Algol evolution in 
very massive binaries. We note that the donor eventually detaches towards 
the end of its main sequence evolution. At core hydrogen exhaustion, the 
donor expands again and initiates the thermal timescale Case\,AB mass 
transfer phase. 

The right panel of Fig.\,\ref{fig:de_reverse} shows the evolution of a 
massive binary model with a 44.6\,$\mso$ donor in a 7.9\,d orbit around 
an accretor of initial mass 22.3\,$\mso$. The donor star (red line) 
fills its Roche lobe (Panel\,`c') at around 4.3\,Myr, initiating thermal 
timescale fast Case\,A mass transfer phase where the mass transfer rate 
rises above 10$^{-3}$\,$M_{\odot}$\,yr$^{-1}$ (Panel\,`b'). During this 
short time, the donor loses $\sim$30\% of its total mass (Panel\,`a'). 
Similar to the semi-detached model, the mass ratio does not invert during 
the fast Case\,A phase, that is, the donor remains the more massive star 
of the binary. 

Unlike the semi-detached model however, this model detaches after the 
fast Case\,A mass transfer phase (see the decline in R/R$_{\rm RL}$ in 
Panel\,`c') and has no nuclear timescale slow Case\,A mass transfer phase. 
The sharp decline in the donor mass increases the luminosity-to-mass ratio 
of the donor. This increases the wind mass-loss rate of the donor 
\citep[][Panel\,`b']{Vink2000}. Due to this increased wind mass-loss rate the radius of 
the donor shrinks upon mass loss \citep{petrovic2005} instead of the usual 
increase in stellar radius during the main sequence. The increased wind 
mass-loss rate may also enable the donor to develop an optically thick 
wind and be observable as a hydrogen-rich WR star. 

For both the example models, the initial mass of the accretors are half 
of that of the donors. Hence, the wind mass-loss rate of the accretors are 
much lower compared to the donors, such that the accretors hardly decrease 
in mass after the fast Case\,A mass transfer phase. Due to the very high 
mass of the binary components, the orbital velocities are of the order of 
100\,km\,s$^{-1}$ (Panel\,`e'), even though the orbital period of the 
semi-detached model is around 100\,d (Panel\,`d'). This may facilitate 
the detection of these reverse Algol systems, despite some of them having 
a high orbital period. 

Figure\,\ref{fig:kipp} shows the evolution of the internal structure 
of the donor and accretor of the binary model that detaches after the 
fast Case\,A mass transfer. We see the retreating convective core of 
the donor (left panel) during the main sequence. The fast Case\,A mass 
transfer removes the outer envelope of the star up to the initial extent 
of the convective core. The convective core of the donor hardly reacts 
to the mass loss, decreasing its mass only slightly. After the fast 
Case\,A mass transfer, the star loses mass faster via wind mass-loss 
than before since the wind mass-loss rate has increased (see Fig.\,\ref{fig:de_reverse}). 
The convective core of the donor continues to recede as well. 

The convective core of the accretor (right panel) remains 
almost constant in mass. Since the accretor is only half as massive 
as the donor, the main sequence lifetime of the accretor is much 
greater than that of the donor, due to which the recession of its 
convective core is much slower compared to that of the donor. The 
accretor in this model reaches critical rotation soon after accreting 
very little mass from the donor (\citealp{packet1981}, see also 
\citealp[][Fig.\,9]{Sen2021b}), because the tidal forces are not 
strong enough to halt its spin-up. Accordingly, our mass transfer 
prescription leads to a low mass transfer efficiency. Since the 
mass transfer efficiency is low, the accretor only gains around 
$\sim$2\,$\mso$ during the fast Case\,A mass transfer. 

\subsection{Surface abundances}

The nitrogen mass fraction inside the convective core of a massive 
star becomes equal to the CNO equilibrium value soon after the onset 
of hydrogen burning. Panels\,`g' and\,`h' of Fig.\,\ref{fig:sd_reverse} 
show that the the surface nitrogen mass fraction enhancement of the 
donors becomes equal to the CNO equilibrium value for the LMC while 
the surface helium mass fraction is only slightly increased during 
fast Case\,A mass transfer. This implies that the fast Case\,A mass 
transfer phase removes the envelope of the donors up to the depth 
where the convective core had developed at the beginning of its main 
sequence. 

During the slow Case\,A mass transfer phase, the wind mass-loss 
removes material from the hydrogen-helium gradient region that forms 
due to the recession of the convective core during the main sequence. 
Layers that are increasingly enriched in helium appears successively 
at the surface of the donor. It is only during the Case\,AB mass transfer 
phase that the envelope stripping reaches the very deep layers of the 
donor where the helium mass fraction is $\sim$0.75. Since the mass 
transfer efficiency in both the models is low, the surface helium and 
nitrogen mass fraction of the accretor does hardly increase from their 
initial values. Moreover, efficient thermohaline mixing quickly mixes 
the helium and nitrogen enriched material from the donor throughout 
the envelope of the accretor. 


\subsection{Orbital period evolution}
\label{subsection_period}

We note that the wind mass-loss rate of the donor (dotted red line in Panel\,`b' 
of Fig\,\ref{fig:de_reverse}) exceeds the nuclear timescale mass transfer 
rate. This is also captured by the evolution of the orbital period of the 
model (Panel\,`d'). Before the onset of fast Case\,A mass transfer, the 
orbital period of the models gradually increased from 100\,d to 120\,d 
(\textit{left panel}) and from 7.9\,d to 8.6\,d (\textit{right panel}), 
primarily due to the wind mass-loss from the donor. Then, during the 
fast Case\,A mass transfer phase, the orbital period decreases rapidly 
because the binary evolves towards equal masses of both components 
\citep{wellstein2001}. However, after the fast Case\,A mass transfer phase, 
the orbital period starts increasing again, despite the binary evolving 
towards a mass ratio of unity. This is because the primary source of mass 
loss from the binary is through the stellar wind of the mass donor and it 
dominates over the effect of nuclear timescale mass transfer. 

In contrast, conventional Algols have a positive orbital period derivative 
after the fast Case\,A phase \citep{Sen2021b} because the mass ratio of the 
binary has already inverted \citep{wellstein2001} and mass transfer is 
occurring at the nuclear timescale and the wind mass-loss rate is lower 
than the nuclear timescale mass transfer rate \citep[see Fig.\,2 of][]{Sen2021b}. 
Hence, we expect the ratio of orbital period to its derivative for Algol 
binaries to capture the nuclear timescale of the donor \citep{Sen2021b}. 
For the case of the reverse Algols however, we expect this ratio to capture 
the wind mass-loss timescale of the donor, which is smaller than its 
nuclear timescale. 

\begin{figure*}
    \centering
    \includegraphics[width=0.48\hsize]{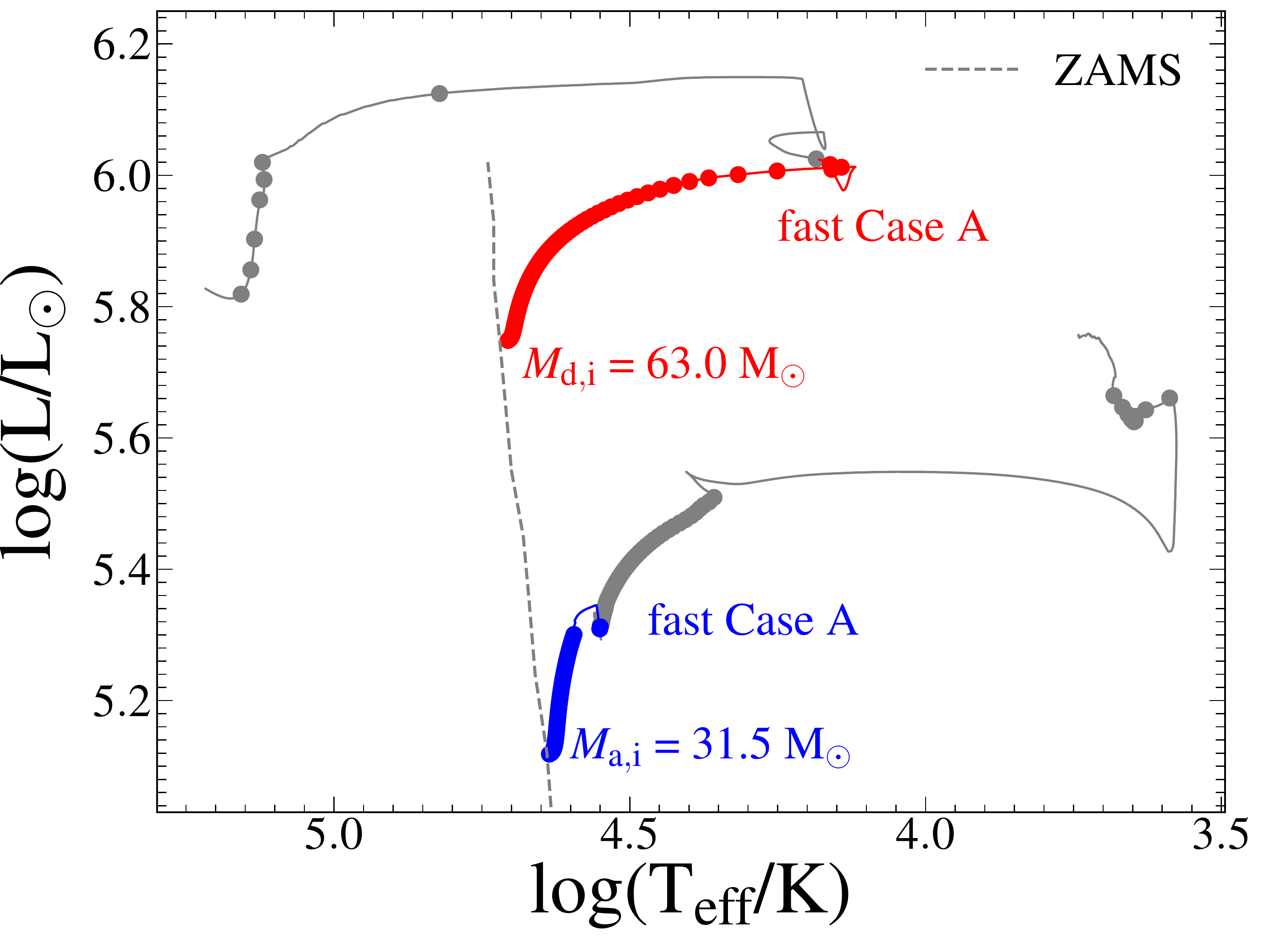}
    \includegraphics[width=0.48\hsize]{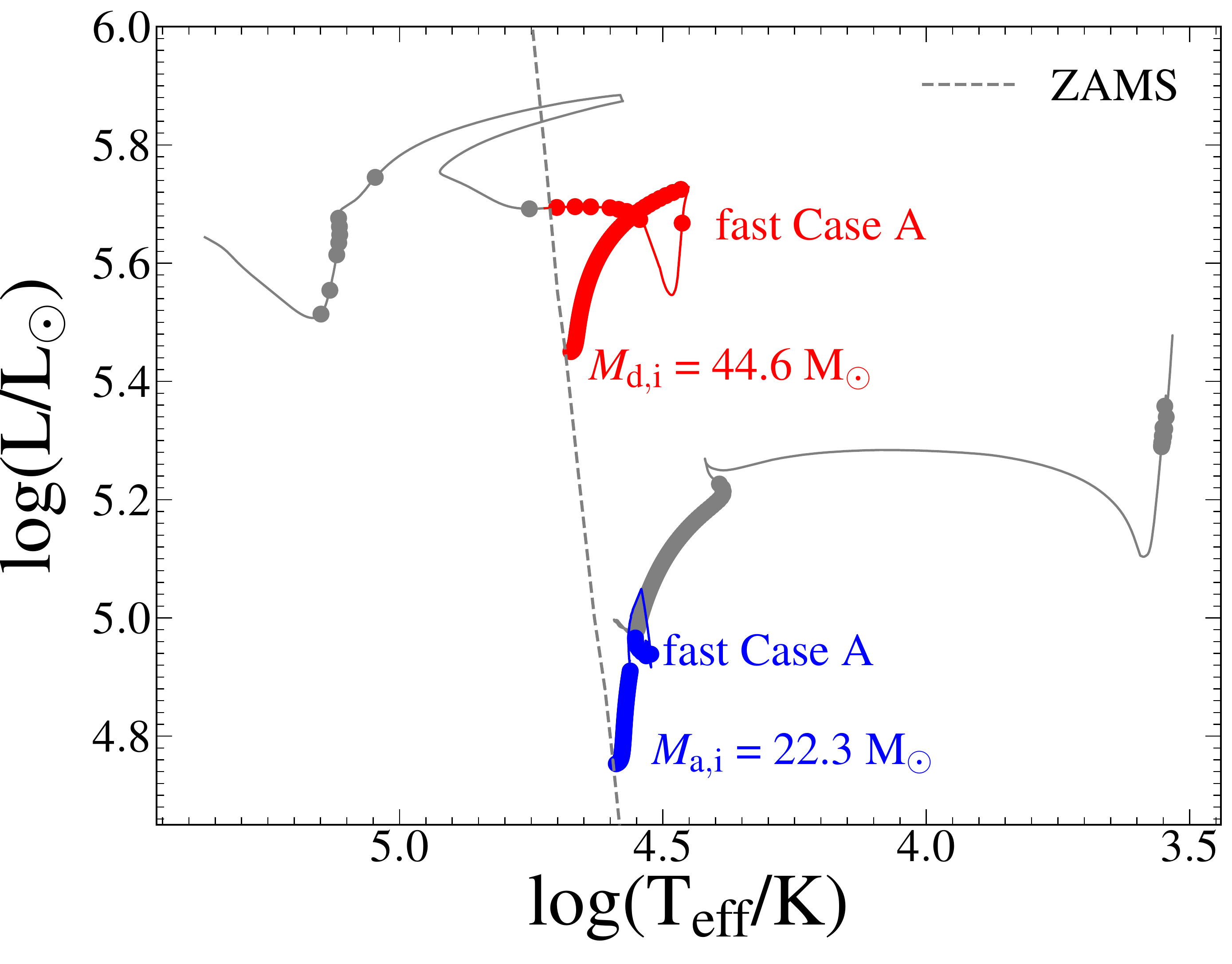}
    \caption{Same as Fig.\,\ref{fig:hrd_BAT99_113} but for two of the exemplary models discussed in Appendix\,\ref{section_examples}. 
    \textit{Left panel:} binary model with an initial orbital period of 100\,d, which experiences a semi-detached reverse Algol phase. \textit{Right panel:} Model with an initial orbital period of 7.9\,d, which experiences a detached reverse Algol phase. See also., Fig.\,\ref{fig:sd_reverse}.}
    \label{fig:hrds}
\end{figure*}

\subsection{Evolution in the Hertzsprung-Russell diagram}
\label{subsection_HRDs}

Figure\,\ref{fig:hrds} show the luminosity and surface temperature of the 
stellar components in the two binary models. In both models, the donors are 
3-6 times more luminous than the accretors at the beginning of core hydrogen 
burning. During the main sequence evolution before the interaction, the 
more massive donor (left panel) reaches lower effective temperatures 
than the less massive donor (right panel), because the envelope 
of the more massive donor has a larger Eddington factor. The luminosity 
of the donors does not decrease significantly during the fast Case\,A mass 
transfer phase \citep[cf. Fig.\,1 of][]{wellstein2001}, with the dip in 
luminosity being smaller for the more massive donor. This is because of 
the decreasing amount of envelope mass that is removed by the Case\,A 
mass transfer phase with increase in mass of the donor. 

Notably, the donors remain more luminous than the accretors after the fast 
Case\,A mass transfer phase for the remainder of their main sequence evolution. 
This is in contrast to the inversion of the luminosity ratio seen in lower 
mass Algol binaries, where the accretors are both more massive and more 
luminous than the donors during the slow Case\,A mass transfer phase \citep{wellstein2001}. 
Also, the more massive donor stays cooler than its accretor as well as the 
less massive donor after the fast Case\,A mass transfer phase. In fact, 
the more massive donor remains within the main sequence band while the 
less massive donor get hot enough to migrate to the left of the main 
sequence towards the end of its core hydrogen burning phase. Since the 
mass transfer efficiency in both the models is low, the mass accretors 
evolve similar to a single star of their corresponding mass, once they 
regains thermal equilibrium after the fast Case\,A mass transfer. 

The mass donors, after undergoing the thermal timescale fast Case\,AB 
mass transfer, get stripped of most of their remaining hydrogen envelope 
\citep{Sen2021b}. They become much hotter than the main sequence stars 
and spend their remaining lifetime on the left of the ZAMS line (denoted 
in grey). The mass accretors on the other hand, evolve similar to single 
stars and spend their post main sequence lifetime as red supergiants. 


\section{Two more Case\,A mass transfer scenarios}
\label{appendix:more_examples}

\begin{figure*}
    \centering
    \includegraphics[width=0.48\linewidth]{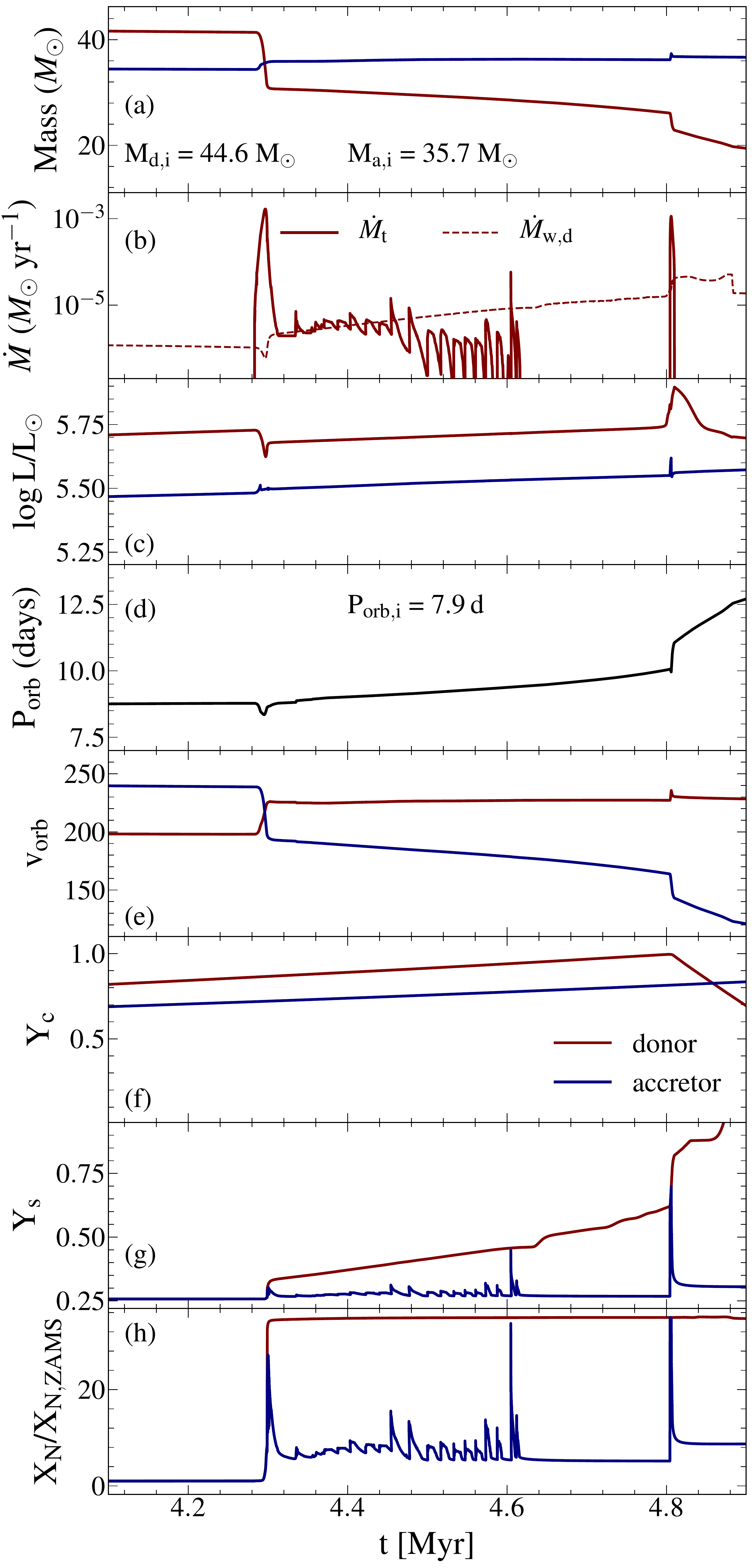}
    \includegraphics[width=0.48\linewidth]{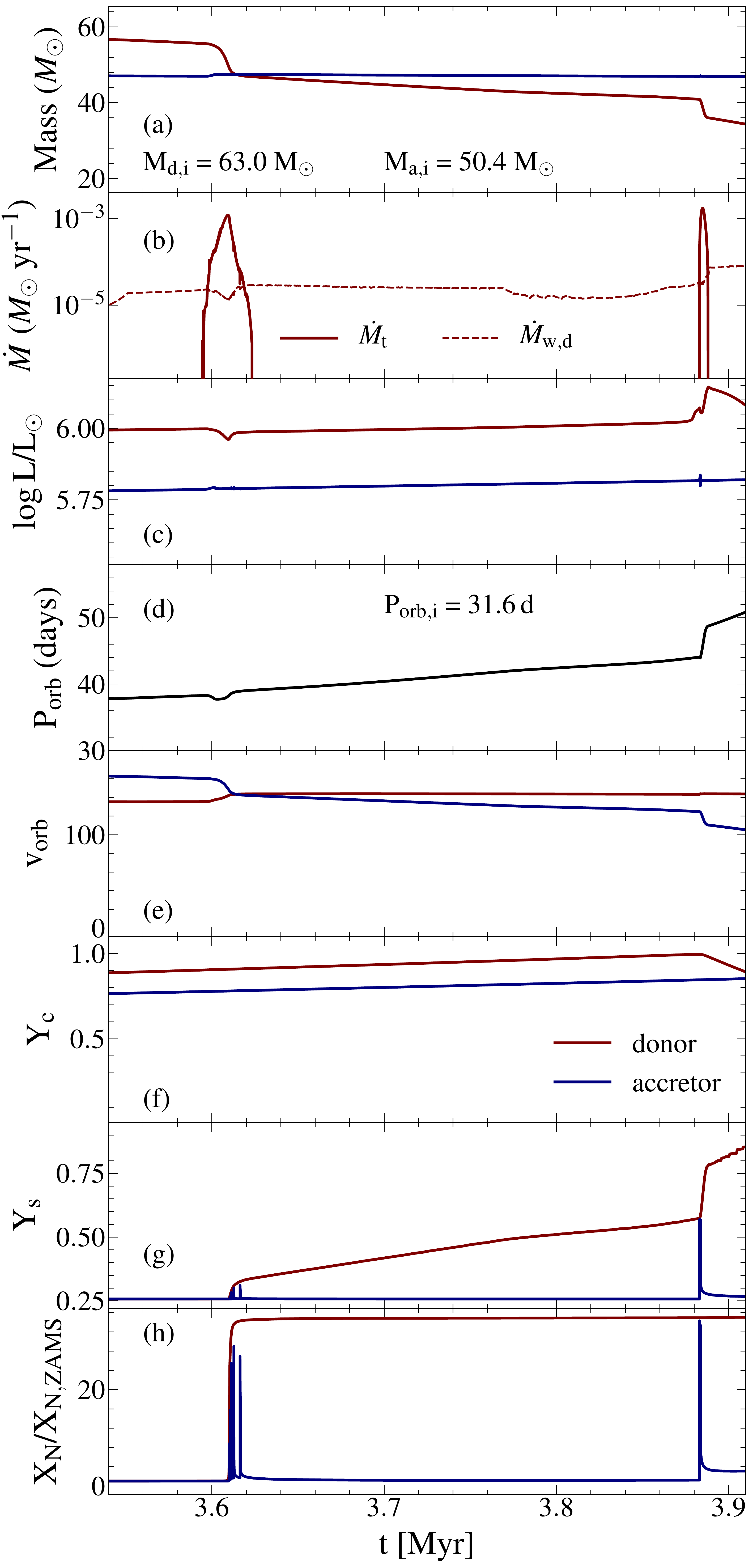}
    \caption{Similar to Fig.\,\ref{fig:sd_reverse} but with Panel\,`c' showing the luminosity of the donor and the accretor. \textit{Left panel:} Initial donor mass, mass ratio and orbital period of the model are 44.6\,$M_{\odot}$, 0.8 and $\sim$7.9\,d respectively. \textit{Right panel:} Initial donor mass, mass ratio and orbital period of the model are 63.0\,$M_{\odot}$, 0.8 and $\sim$31.6\,d respectively. }
    \label{fig:sd_algol}
    \label{fig:de_normal}
\end{figure*}



Figure\,\ref{fig:sd_algol} shows two more examples 
of binary models where the mass ratio does get inverted after the fast 
Case\,A mass transfer phase (Panel\,`a') but the donors are so overluminous 
for their mass that their absolute luminosity remains higher than the 
accretor (Panel\,`c'). We see that the model with an initial donor mass 
of 44.6\,$\mso$ goes through a classical Algol phase but the luminosity 
of the donor does not dip significantly after the fast Case\,A mass 
transfer phase \citep[cf. Fig.\,1 of][]{wellstein2001}. This leads to 
another unique configuration in an Algol binary where the less massive 
donor is more luminous than the accretor. We also see that the model 
with an initial donor mass of 63\,$\mso$ detaches after the fast Case\,A 
mass transfer. This leads to a peculiar configuration in a detached binary 
where the less massive star is more luminous than the more massive star, 
while on the main sequence band in the HR diagram.

\end{appendix}

\end{document}